\documentclass[12pt,english]{article}
\usepackage{graphicx,array}
\pdfoutput=1
\usepackage{cite}
\usepackage{color}
\usepackage{latexsym}
\usepackage{amsthm}
\usepackage{amsmath}
\usepackage[titletoc]{appendix}
\usepackage{enumitem}
\usepackage{amssymb}
\usepackage{hyperref}
\usepackage[hang,flushmargin]{footmisc} 

\def\simgt{\mathrel{\lower2.5pt\vbox{\lineskip=0pt\baselineskip=0pt
           \hbox{$>$}\hbox{$\sim$}}}}
\def\simlt{\mathrel{\lower2.5pt\vbox{\lineskip=0pt\baselineskip=0pt
           \hbox{$<$}\hbox{$\sim$}}}}

\newcommand{\eq}[1]{\begin{align}#1\end{align}}
\newcommand{\Eq}[1]{Eq.~(\ref{#1})}

\newcommand{\Sec}[1]{Sec.~\ref{#1}}
\newcommand{\Secs}[2]{Secs.~\ref{#1} and \ref{#2}}
\newcommand{\Fig}[1]{Fig.~\ref{#1}}

\newcommand{\Ref}[1]{Ref.~\cite{#1}}

\newcommand{\mPl}{m_{\rm Pl}}

\newcommand{\Mink}{\textrm{Mink}}
\newcommand{\Milne}{\textrm{Milne}}
\newcommand{\Rind}{\textrm{Rind}}
\newcommand{\AdS}{\textrm{AdS}}
\newcommand{\dAdS}{\partial\textrm{AdS}}
\newcommand{\ddS}{\partial\textrm{dS}}
\newcommand{\dS}{\textrm{dS}}
\newcommand{\CFT}{\textrm{CFT}}
\newcommand{\R}{\mathbb{R}}

\newcommand*\oline[1]{%
  \vbox{%
    \hrule height 0.5pt
    \kern0.68ex
    \hbox{%
      \kern-0.1em
      \ifmmode#1\else\ensuremath{#1}\fi
      \kern-0.1em
    }
  }
}

\definecolor{nicered}{rgb}{0.7,0.1,0.1}
\definecolor{nicegreen}{rgb}{0.1,0.5,0.1}
\hypersetup{colorlinks=true,citecolor=blue,linkcolor=blue,urlcolor=blue}
\usepackage[tracking=true]{microtype}
\usepackage[T1]{fontenc}

\usepackage{fancyhdr}
\fancypagestyle{empty}{
	\setlength{\headheight}{15.2pt}
	\fancyhf{}
	 
	\fancyhead[R]{CALT-TH-2016-024 \\ UMD-PP-017-010} 
}

\begin{document}
\interfootnotelinepenalty=10000
\baselineskip=18pt
\hfill
\hfill 

\vspace{1cm}
\thispagestyle{empty}
\begin{center}
{\Large \bf
4D Scattering Amplitudes and    \\  \medskip Asymptotic Symmetries from 2D CFT}\\
\bigskip\vspace{.5cm}{
{\large Clifford Cheung$^1$, Anton de la Fuente$^2$, and Raman Sundrum$^2$}
} \vspace{.5cm}

 {$^1$\it Walter Burke Institute for Theoretical Physics, \\
    California Institute of Technology, Pasadena, CA 91125}\\
{$^2$\it Department of Physics, University of Maryland, College Park, MD 20742}
    \let\thefootnote\relax\footnote{e-mail: \url{clifford.cheung@caltech.edu}, ~
     \url{delafuea@umd.edu}, ~
      \url{raman@umd.edu}} \\
 \end{center}
\bigskip
\centerline{\large\bf Abstract}
\bigskip

We reformulate the scattering amplitudes of 4D flat space gauge theory and gravity in the language of a 2D CFT on the celestial sphere.  
The resulting CFT structure
exhibits an OPE constructed from 4D collinear singularities, as well as infinite-dimensional Kac-Moody and Virasoro algebras encoding the asymptotic symmetries of 4D flat space.
We derive these results by recasting 4D dynamics in terms of a convenient foliation of flat space into 3D Euclidean $\AdS$ and Lorentzian $\dS$ geometries.  Tree-level scattering amplitudes take the form of Witten diagrams for a continuum of $\textrm{(A)dS}$ modes, which are in turn equivalent to CFT correlators  via the (A)dS/CFT dictionary.  
The Ward identities for the 2D conserved currents are dual to 4D soft theorems, while the bulk-boundary propagators of massless (A)dS modes are superpositions of the leading and subleading Weinberg soft factors of gauge theory and gravity.  In general, the massless (A)dS modes are 3D Chern-Simons gauge fields describing the soft, single helicity sectors of 4D gauge theory and gravity.  
Consistent with the topological nature of Chern-Simons theory,  Aharonov-Bohm effects record the ``tracks'' of hard particles in the soft radiation, leading to a simple characterization of gauge and gravitational memories.    Soft particle exchanges between hard processes define the Kac-Moody level and Virasoro central charge, which are thereby related to the 4D gauge coupling and gravitational strength in units of an infrared cutoff.  Finally, we discuss a toy model for black hole horizons via a restriction to the Rindler region.

\setcounter{footnote}{0}

\newpage
\tableofcontents

\newpage

\section{Introduction}

\label{sec:intro}

The AdS/CFT correspondence \cite{maldacenaAdSCFT,polyakovAdSCFT,wittenAdSCFT,magoo,polchinskiTASI,ramanAdSCFT,penedonesTASI} has revealed profound insights into the dualities equating theories with and without gravity.   As an explicit formalism, it has also given teeth to the powerful notion of holography, fueling concrete progress on longstanding puzzles in an array of subjects, ranging from black hole physics to strongly coupled dynamics.  Still, AdS/CFT professes the limits of its own applicability: the entire construction rests pivotally on the infrastructure of warped geometry. 

In this paper, we explore a potential strategy for channeling the  power of AdS/CFT  into 4D Minkowski spacetime. 
 This ambitious goal has a long history \cite{polchinskiFlatLimit,susskindFlatLimit,penedonesGiddingsFlatLimit,penedonesMellin,jaredLiamFlatLimit,jaredLiamAnalyticity,jaredLiamUnitarity,penedonesNew1,penedonesNew2}, typically with a focus on AdS/CFT in the limit of infinite AdS radius.  Here we follow a different path, in line with the seminal work of\cite{deBoerSolodukhin,solodukhin}. The crux of our approach is to foliate Minkowski spacetime into a family of warped 3D slices for which the methodology of AdS/CFT is applicable,   recasting the dynamics of 4D flat space into the grammar of a 2D CFT.\footnote{See \cite{bigYellowBook} and references therein for a handy review of 2D CFT.}
We derive the central objects of this conjectured 2D CFT---namely the conserved currents and stress tensor---and show how the corresponding Kac-Moody and Virasoro algebras
  beautifully encode the asymptotic symmetries of 4D gauge theory \cite{stromingerFirstPaper,stromingerQED,stromingerKacMoody,stromingerMassiveQED,stromingerMagnetic} and gravity \cite{bms1,bms2,belgiansSuperrotations}. Our results give a unified explanation for the deep connections recently discovered\cite{stromingerFirstPaper,stromingerBMS1,stromingerBMS2,stromingerVirasoro,stromingerQED,stromingerQEDsubleading,stromingerKacMoody,stromingerMassiveQED,stromingerMagnetic,stromingerFermionic} between asymptotic symmetries and 4D soft theorems
  \cite{weinberg,low,burnettKroll, White, cachazoStrominger}, allowing us to extend and understand these results further. 
  As we will see, the 2D current algebras are dual to 3D Chern-Simons (CS) gauge fields that describe soft fields in 4D, and for which the phenomena of gauge \cite{qedMemory,pasterskiQEDmemory,susskindQEDmemory} and gravitational ``memories'' \cite{memoryOriginal1,memoryOriginal2,memoryOriginal3,stromingerBMSmemory,stromingerVirasoroMemory} take the form of abelian and non-abelian Aharonov-Bohm effects \cite{aharonovBohm,nonAbelianAB1,nonAbelianAB2}.

 Let us now discuss our results in more detail.  In \Sec{sec:bulkcoordinates},  we set the stage by defining a convenient set of coordinates for 4D Minkowski spacetime ($\Mink_4$).  These coordinates are formally anchored to a fixed origin  \cite{deBoerSolodukhin,solodukhin,stewartSchwartz,campiglia,Costa:2012fm} intuitively representing the location of a hard scattering process. In turn, this choice naturally divides $\Mink_4$ into two regions: the 4D Milne spacetimes ($\Milne_4$) past and future time-like separated from the origin, and the 4D spherical Rindler spacetime ($\Rind_4$)  space-like separated from the origin.   We then choose coordinates in which $\Milne_4$ and $\Rind_4$ are foliated into slices at a fixed proper distance from the origin, or equivalently at fixed Milne time and Rindler radius, respectively.  Each Milne slice is equivalent to 3D Euclidean anti-de Sitter space ($\AdS_3$).  While this geometry is purely spatial from the 4D viewpoint, we will for notational convenience refer to it as $\AdS_3$ with the Euclidean signature implied.  Similarly, each Rindler slice is equivalent to Lorentzian de Sitter ($\dS_3$) spacetime.   
 
\begin{figure}[t]
\begin{center}
\includegraphics[width=0.95\textwidth]{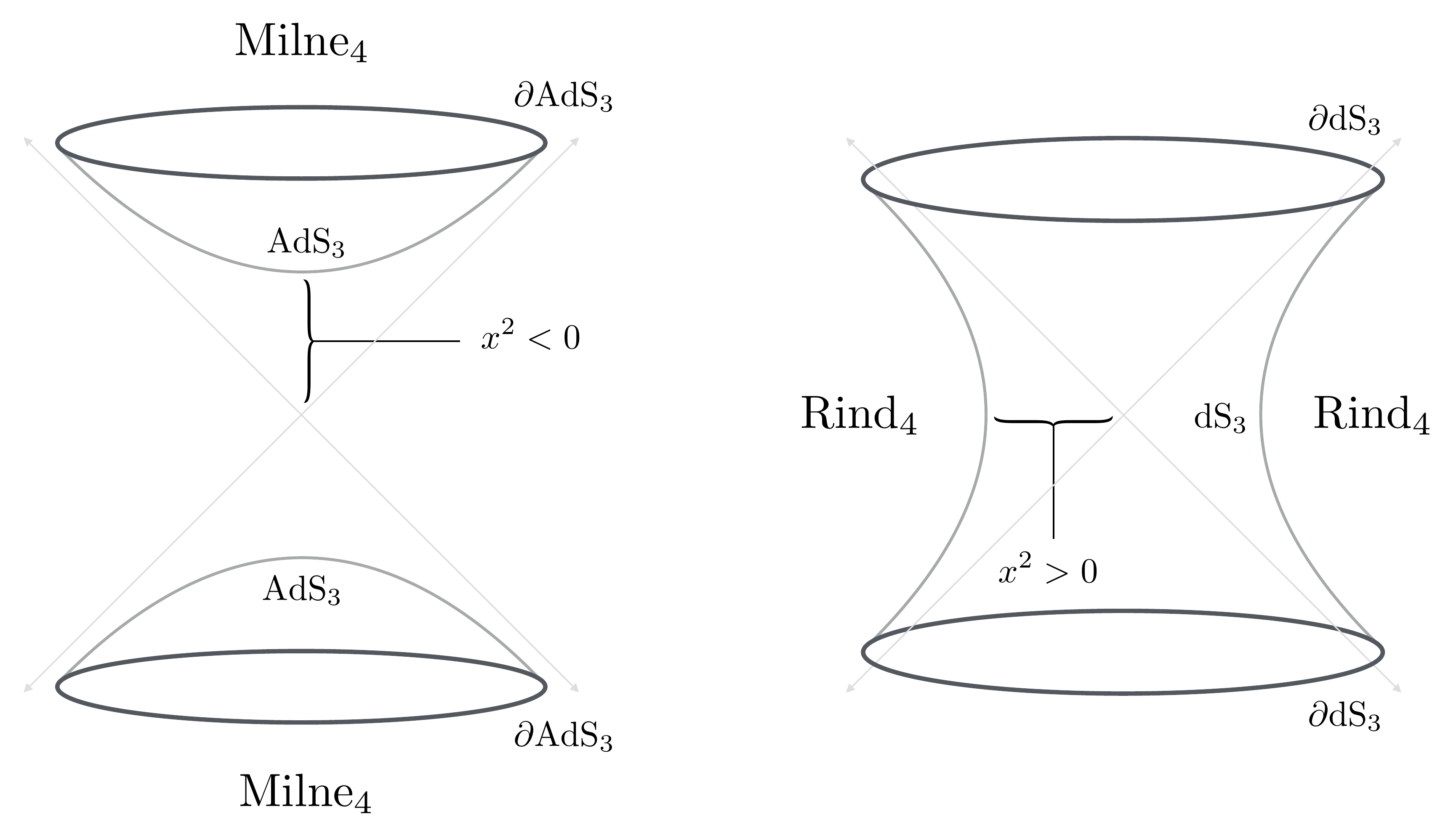}
\end{center}
\vspace*{-0.5cm}
\caption{Minkowski space is divided into Milne and Rindler regions which are time-like and space-like separated from the origin, respectively.  Each region is then foliated into a family of warped slices, each at a fixed proper distance from the origin.     }
 \label{fig:spacetime}
\end{figure}

  In \Sec{sec:boundarycoordinates}, we show how the corresponding $\AdS_3$ and $\dS_3$ boundaries ($\dAdS_3$ and $\ddS_3$) define a 2D celestial sphere at null infinity---the natural
  home of  massless asymptotic states.  
        By choosing the analog of Poincare patch coordinates on the warped slices, we find that the celestial sphere is labeled by complex variables $(z,\bar z)$ that coincide with the projective spinor helicity variables frequently used in the study of scattering amplitudes.   The geometry of our setup is depicted in \Fig{fig:spacetime}, and our basic approach is outlined in \Sec{sec:approach}.

Armed with a foliation of $\Milne_4$ into $\AdS_3$ slices, we apply the $\AdS_3/\CFT_2$ dictionary, bearing in mind that the underlying spacetime is actually flat
 \cite{deBoerSolodukhin,solodukhin}.  To do so, in \Secs{sec:gaugemodeexpansion}{sec:scalingdimensions} we  apply separation of variables  to 
  decompose all the degrees of freedom in $\Milne_4$ into 
  ``harmonics'' in Milne time, yielding a continuous spectrum of ``massive'' $\AdS_3$ fields. Here the $\AdS_3$ ``mass'' of each field is simply its Milne energy.\footnote{This energy is in general not conserved in the ``expanding Universe'' defined by Milne spacetime, but it will be in a number of Weyl invariant theories of interest.}   In \Sec{sec:gaugewittendiagrams} we go on to show that the Witten diagrams of $\AdS_3$ fields are precisely equal to flat space scattering amplitudes in $\Milne_4$, albeit with a modified prescription for LSZ reduction substituting $\AdS_3$ bulk-boundary propagators for plane waves. 
    In turn, the $\AdS_3/\CFT_2$ correspondence offers a formalism to recast these scattering amplitudes as correlators of a certain $\CFT_2$  living on the celestial sphere.   The operator product expansion corresponds to singularities in $(z,\bar z)$ arising from collinear limits in the angular directions.
    
    In \Sec{sec:continuation}, we show how the $\AdS_3/\CFT_2$ dictionary in $\Milne_4$ dovetails with the 
  $\dS_3/\CFT_2$ dictionary \cite{stromingerdSCFT,wittenQGindS,maldacenaNonGaussianities,stromingerHigherSpin} in $\Rind_4$  by analytic continuation through the ambient $\Mink_4$ embedding space.   Here the mechanics of this continuation, as well as our calculations in general, are greatly simplified by employing the elegant embedding formalism of \cite{dirac,penedonesDIS,weinbergEmbedding,penedonesRychkovPoland,penedones}.  Notably, the appearance of $\dS_3$ suggests that the underlying $\CFT_2$ is {\it non-unitary}, as we see in detail.  Putting it all together in \Sec{sec:MAdSCFT}, we are then able to extend the mapping between 4D scattering amplitudes and 2D correlators to all of Minkowski spacetime.     
     
A natural question now arises: which 4D scattering amplitudes are dual to the 2D correlators of conserved currents? For scattering amplitudes in the Milne region, the Witten diagrams for these correlators will involve massless 
$\AdS_3$ fields.  According to our decomposition into Milne harmonics, these massless modes have vanishing Milne energy, and thus correspond to the Milne soft limit of particles in the 4D scattering amplitude.  
 In the case of gauge theory, we show in \Sec{sec:conservedcurrents} that the Milne soft limit coincides precisely with the usual soft limit taken with respect to Minkowski energy.  As a result, the Ward identity for a conserved current in 2D is literally equal to the leading Weinberg soft theorem for gauge bosons in 4D, which we show explicitly for abelian gauge theory with matter as well as Yang-Mills (YM) theory.  We thereby conclude that the conserved currents of the $\CFT_2$ are dual to soft gauge bosons in $\Mink_4$.  It is attractive that the $\AdS_3/\CFT_2$ dictionary automatically guides us to identify 4D soft limits with 2D conserved currents.  Afterwards, in \Sec{sec:KM} we show how the existence of a 2D holomorphic conserved current relates to the presence of an infinite-dimensional Kac-Moody algebra.\footnote{Such a structure was observed long ago in amplitudes \cite{Nair}, serving as inspiration for the twistor string\cite{wittenTwistorString}.}

Next, we go on to construct the explicit  $\AdS_3$ dual of the $\CFT_2$ for the current algebra subsector.  In \Sec{sec:gaugeCS}, we show that soft gauge bosons of a single helicity comprise a 3D topological CS gauge theory in $\AdS_3$ whose dual is the 2D chiral Wess-Zumino-Witten (WZW) model \cite{wittenJones,seibergMoore,wittenHolomorphic,stromingerCSAdSCFT} discussed in \Sec{sec:WZW}.  As is well-known, this theory is a 2D CFT imbued with an infinite-dimensional Kac-Moody algebra.  We show explicitly how hard particles in 4D  decompose into massive 3D matter fields that source the CS gauge fields.   Afterwards, we discuss the Kac-Moody level $k_{\rm CS}$ and its connection to internal exchange of soft gauge bosons.  Our results suggest that the level is related to the 4D YM gauge coupling via $k_{\rm CS} \sim 1/g_{\rm YM}^2$.

We also show in \Sec{sec:gaugememory} how the topological nature of CS theories reflects the remarkable phenomenon of 4D gauge  ``memory'' \cite{qedMemory,pasterskiQEDmemory,susskindQEDmemory} in which soft fields record the passage of hard particles carrying conserved charges through specific angular regions on the celestial sphere. In our formulation, these memory effects are naturally encoded as abelian and non-abelian Aharonov-Bohm phases from the encircling of hard particle ``tracks'' by CS gauge fields.

Interestingly, Ref.~\cite{softHair} proposed that gauge and gravitational memories have the potential to encode copious ``soft hair'' on black hole horizons, offering new avenues for understanding the information paradox, as reviewed in \cite{harlow}.  While black hole physics is not the primary focus of this work, our formalism does give a natural framework to study a toy model for black hole horizons which we present in \Sec{sec:toymodel}.  In particular, by excising the Milne regions of spacetime, we are left with a Rindler spacetime that describes a family of radially accelerating observers.  We find that the $\CFT_2$ structure extends to include the early and late time wavefunction at the Rindler horizon.  In particular, the 2D conserved currents are dual to CS soft fields that record the insertion points of hard particles that puncture the horizon and that escape to null infinity.

       In a parallel analysis for gravity, we show in \Sec{sec:stresstensor} that the Ward identity for the 2D stress tensor is an angular  convolution of the subleading Weinberg soft theorem for gravitons in 4D.  
As for any $\CFT_2$, this theory is equipped with an infinite-dimensional Virasoro algebra that we discuss in \Sec{sec:virasoro}. Since the global $SL(2,\mathbb{C})$ subgroup is nothing but the 4D Lorentz group, these Virasoro symmetries are aptly identified as the ``super-rotations'' of the extended BMS algebra of asymptotic symmetries in 4D flat space \cite{bms1,bms2,belgiansSuperrotations}.   We then consider the case of subleading soft gravitons and the $\CFT_2$ stress tensor in \Sec{sec:gravityCS}, arguing that the dual theory is simply $\AdS_3$ gravity, which famously is equivalent to a CS theory in 3D \cite{witten3dGravity1,witten3dGravity2}.   Afterwards, we go on to discuss the connections between 4D gravitational memory, and the Virasoro algebra.  While the value of the Virasoro central charge $c$ is subtle, our physical picture suggests that $c \sim \mPl^2 L_{\rm IR}^2$, where $\mPl$ is the 4D Planck scale and $L_{\rm IR}$ is an infrared cutoff.   We then utilize the extended BMS algebra \cite{belgiansChargeAlgebra} to derive the $\CFT_2$ Ward identity associated with ``super-translations'' \cite{bms1,bms2}, and we confirm that they correspond to the leading Weinberg soft theorem for gravitons \cite{stromingerBMS1,stromingerBMS2}. 

Finally, let us pause to orient our results within the grander ambitions of constructing a holographic dual to flat space.   Our central results rely crucially on the soft limit in 4D, wherein lie the hallmarks of 2D CFT.  At the same time, a holographic dual to flat space will necessarily describe all 4D dynamics, including the soft regime.  Hence, our results imply that the soft limit of any such dual will be described by a CFT.  In this sense, the CFT structure derived in this paper should be interpreted as a stringent constraint on any holographic dual to flat space.

{\it Note added: during the final stages of preparation for this paper, \Ref{NewStrominger} appeared, also deriving a 2D stress tensor for 4D single soft graviton emission.}

\section{Setup}

\label{sec:coords}
As outlined in the introduction, our essential strategy is to import the holographic correspondence into flat space by reinterpreting $\Mink_4$ as the embedding space for a family of $\AdS_3$ slices \cite{deBoerSolodukhin,solodukhin}.  To accomplish this, we foliate $\Mink_4$ into a set of warped geometries and mechanically invoke the $\AdS_3/\CFT_2$ dictionary, recasting its implications as old and new facts about flat space scattering amplitudes.  We now define bulk and boundary coordinates natural to achieve this mapping.

\subsection{Bulk Coordinates}
\label{sec:bulkcoordinates}

To begin, we define 4D Cartesian coordinates $x^\mu = (x^0,x^1,x^2,x^3)$ associated with the flat metric,
\eq{
  ds^2_{\Mink_4} = \eta_{\mu\nu} dx^\mu dx^\nu ,
}
and labeled by Greek indices $(\mu,\nu,\ldots)$ hereafter.  As outlined in the introduction, it will be convenient to organize spacetime points in Minkowski space according to their proper distance from the origin.  This partitions flat space into Milne and Rindler regions that are time-like and space-like separated from the origin.  

\subsubsection{Milne Region}

We foliate the 4D Milne region into hyperbolic slices of a fixed proper distance from the origin,
\eq{
x^2= - e^{2\tau},
\label{eq:tau_def}
}
where $\tau$ is the Milne time coordinate.  Together with the remaining spatial directions, $\tau$ defines a set of 4D hyperbolic Milne coordinates,
\eq{
Y^I = (\tau,\rho,z,\bar z),
}
denoted by upper-case Latin indices $(I,J,\ldots)$ hereafter. The Milne coordinates $Y^I$ are related to the Cartesian coordinates $x^\mu$ according to
\eq{
x^0 &=  \frac{e^\tau  \rho}{2}\left(1+\frac{1}{\rho^2}(1+z\bar z)\right),\qquad\qquad   x^1+ix^2 =  \frac{e^\tau z}{\rho},\nonumber\\
x^3 &=  \frac{e^\tau  \rho}{2}\left(1-\frac{1}{\rho^2}(1-z \bar z)\right),\qquad\qquad x^1-ix^2 =  \frac{e^\tau \bar z}{\rho} .
\label{eq:coord_transform}
}
The domain for each Milne coordinate is $\tau ,\rho \in\mathbb{R}$ and $z,\bar z\in \mathbb{C}$.  The regions $\rho >0 $ and $\rho<0$ correspond to the two halves of  $\Milne_4$---that is, the future and past Milne regions circumscribed by the future and past lightcones of the origin, respectively.  So depending on the sign of $\rho$, the $\tau \rightarrow +\infty$ limit corresponds to either the asymptotic past or the asymptotic future.  On the other hand, the $\tau\rightarrow -\infty$ limit corresponds to the $x^2=0$ boundary dividing the Milne and Rindler regions.  In the context of a standalone Rindler spacetime, this boundary is known as the Rindler horizon.\footnote{More precisely, we are considering a spherical rather than the standard planar Rindler region reviewed in \cite{birrellDavies}.}  In the current setup, however, this horizon is a coordinate artifact simply because the underlying Minkowski space seamlessly joins the Milne and Rindler regions.  Last but not least, $(z,\bar z)$ denote complex stereographic coordinates on the celestial sphere.  Note that the physical angles on the sky labeled by $(z,\bar z)$ are antipodally identified for $\rho >0$ and $\rho<0$, due to the diametric mapping between celestial spheres in the asymptotic past and the asymptotic future.

By construction, the Milne coordinates are defined so that $\Milne_4$ decomposes into a family of Euclidean $\AdS_3$ geometries,\eq{
ds^2_{\Milne_4} &=  G_{IJ}(Y) dY^I dY^J = e^{2\tau} \left(-d\tau^2 + ds^2_{\AdS_3} \right). 
\label{eq:Milne_coords}
}
Each slice at fixed $\tau$ describes a 3D geometry equivalent to  Euclidean $\AdS_3$ spacetime in Poincare patch coordinates \cite{magoo}, so
\eq{
ds^2_{\AdS_3} &= g_{ij}(y) d y^{i} d y^{j} = \frac{1}{\rho^2} (d\rho^2 +dz d\bar z),
\label{eq:hyper_coords}
}
where lower-case Latin indices $(i,j,\ldots)$ denote $\AdS_3$ coordinates,
\eq{
 y^{i} = (\rho,z,\bar z),
\label{eq:hatted_coords}
}
which are simply a restriction of the Milne coordinates, $Y^I = (\tau,y^i)$.

From \Eq{eq:hyper_coords} it is obvious that $\rho$ corresponds to the radial coordinate of $\AdS_3$ and the $\rho\rightarrow 0$ limit defines the boundary  $\dAdS_3$.  Interpolating between the past and future Milne regions corresponds to an analytic continuation of the $\AdS_3$ radius $\rho$ to both positive and negative values. 

\subsubsection{Rindler Region}

A similar analysis applies to the 4D Rindler region, which we foliate with respect to
\eq{
x^2 = e^{2\rho},
\label{eq:Rindler_x2}
}
where $\rho$ is now the Rindler radial coordinate.   Like before, we can define hyperbolic Rindler coordinates, $Y^I=(\rho,\tau,z,\bar z)$, with the associated metric,
\eq{
ds^2_{\Rind_4} &=  G_{IJ}(Y) dY^I dY^J = e^{2\rho} \left(d\rho^2 + ds^2_{\dS_3} \right).
\label{eq:Rindler_metric}
}
Splitting the Rindler coordinates by $Y^I = (\rho, y^i)$, we see that each slice at fixed $\rho$ defines a Lorentzian $\dS_3$ spacetime parameterized by $y^i = (\tau, z, \bar z)$ and the corresponding metric,
\eq{
ds^2_{\dS_3} &= g_{ij}(y) d y^{i} d y^{j} = \frac{1}{\tau^2} (-d\tau^2 +dz d\bar z),
}
where $\tau$ is the conformal time of $\dS_3$.

\subsection{Boundary Coordinates}
\label{sec:boundarycoordinates}

Given a hyperbolic foliation of Minkowski space, it is then natural to consider the spacetime boundary associated with each warped slice.   To be concrete, let us focus here on $\Milne_4$, although a similar story will apply to $\Rind_4$.  

Using the Milne coordinates in \Eq{eq:coord_transform}, we express an arbitrary spacetime point in $\Milne_4$ as
\eq{
x^\mu &=e^\tau\left(\frac{k^\mu}{\rho}  + \rho   q^\mu\right),
\label{eq:x_expand}
}
where we have defined the null vectors,
\eq{
k^\mu = \frac{1}{2} \left( 1+ z \bar z, z+ \bar z, -iz + i \bar z, -1 + z \bar z\right) \qquad \textrm{and} \qquad
q^\mu = \frac{1}{2} \left( 1,0,0,1\right).
\label{eq:nq_def}
}
In terms of the celestial sphere, $k^\mu$ is a vector pointing in the $(z,\bar z)$ direction while $q^\mu$ is a reference vector pointing at  complex infinity.  Of course, while $q^\mu$ describes a certain physical angle on the sky, this is a coordinate artifact without any physical significance.

Given a null vector $k^\mu$ it is natural to define polarization vectors,
\eq{
\epsilon^\mu &=\frac{1}{2} \left(  \bar z, 1, -i, \bar z \right)  \nonumber \\
\bar\epsilon^\mu &= \frac{1}{2} \left( z, 1, i, z\right) ,
\label{eq:eebar_def}
}
where $\epsilon$ and $\bar \epsilon$ correspond to $(+)$ and $(-)$ helicity states, respectively.    As usual, the helicity sum over products of polarization vectors yields a projector onto physical states, 
\eq{
\epsilon^\mu \bar\epsilon^\nu+\epsilon^\nu \bar\epsilon^\mu = \frac{1}{2} \left(  \eta^{\mu\nu}-\frac{  q^\mu k^\nu + q^\nu k^\mu }{qk}\right),
}
where $qk =-1/2$ is actually constant.
Note also that the polarization vectors $\epsilon^\mu$ and $\bar\epsilon^\mu$ and the reference vector $q^\mu$ are compactly expressed in terms derivatives of $k^\mu$, 
\eq{
\epsilon^\mu &=  \partial_z k^\mu \nonumber\\
\bar\epsilon^\mu &= \partial_{\bar z} k^\mu \nonumber\\
q^\mu &= \partial_z \partial_{\bar z} k^\mu .
}
The above expressions will be quite useful for manipulating expressions later on.

To go to the boundary of $\AdS_3$ we take the limit of vanishing radial coordinate, $\rho \rightarrow 0$.  According to \Eq{eq:x_expand}, any spacetime point at the boundary approaches a null vector,
\eq{
x^\mu &\overset{\rho \rightarrow 0}{=} \frac{e^\tau k^\mu}{\rho},
}
so $\dAdS_3$ is the natural arena for describing massless degrees of freedom.   To appreciate the significance of this, recall that the in and out states of a scattering amplitude are inserted in the asymptotic past and future, defined by $\tau \rightarrow + \infty$.  For massless particles, this implies that null trajectories at $\tau\rightarrow +\infty$ should approach $\rho\rightarrow 0$ so that asymptotic states originate at $\dAdS_3$ in the far past or terminate at $\dAdS_3$ in the far future.  Said more precisely, $\dAdS_3$ is none other than past and future null infinity restricted to the Milne region.\footnote{Past and future null infinity in the Rindler region is contained in the boundary of $\dS_3$.}  Hence, $\dAdS_3$ is a natural asymptotic boundary associated with the scattering of massless particles.  

Finally, let us comment on the unexpected connection between our coordinates and the spinor helicity formalism commonly used in the study of scattering amplitudes.  In particular, while the specific form of $k^\mu$ in \Eq{eq:nq_def} was rigidly dictated by the choice of Poincare patch coordinates on $\AdS_3$, it also happens to be that
\eq{
k^\mu  &=\lambda \sigma^\mu \bar\lambda ,
}
where $\lambda$ and $\bar \lambda$ are projective spinors,
\eq{
\lambda = (z,1) \qquad \textrm{and} \qquad \bar \lambda =(\bar z,1),
}
in a normalization where ${\rm tr}(\sigma^\mu \bar\sigma^\nu) = \eta^{\mu\nu}/2$.
Here $\lambda$ and $\bar \lambda$ are defined modulo rescaling, {\it i.e.}~modulo the energy of the associated momentum.  This projective property implies that the only invariant kinematic data stored in $\lambda$ and $\bar \lambda$ is angular. 

Meanwhile, the reference vector $q^\mu$ can also be expressed in spinor helicity form,
\eq{
q^\mu  &=\eta \sigma^\mu \bar\eta,
}
where $\eta$ and $\bar \eta$ are reference spinors,
\eq{
\eta = (1,0) \qquad \textrm{and} \qquad \bar \eta =(1,0),
}
and the polarization vectors take the simple form,
\eq{
\epsilon^\mu &= \eta \sigma^\mu \bar\lambda \nonumber \\
\bar\epsilon^\mu &=\lambda \sigma^\mu \bar\eta .
\label{eq:eebar2_def}
}
Thus, our hyperbolic foliation of Minkowski space has induced a coordinate system on the boundary that coincides  with projective spinor helicity variables in a gauge specified by a particular set of reference spinors.  

As usual, we can combine spinors into Lorentz invariant angle and square brackets,
\eq{
\langle 12\rangle = \lambda_{1\alpha} \lambda_{2\beta} \epsilon^{\alpha\beta} = z_1-z_2  \qquad   \textrm{and} \qquad [12]=\bar\lambda_{1\dot\alpha} \bar\lambda_{2\dot\beta} \epsilon^{\dot\alpha\dot\beta} = \bar z_1 -\bar z_2.
}
Meanwhile, the invariant mass of two null vectors,
\eq{
-(k_1 +k_2)^2 = \langle12 \rangle [12] = |z_1-z_2|^2,
}
is the natural distance between points on the celestial sphere.   

As is familiar from the context of scattering amplitudes, expressions typically undergo drastic simplifications when expressed in terms of spinor helicity variables.  For example, the celebrated Parke-Taylor formula for the color-stripped MHV amplitude in non-abelian gauge theory is
\eq{
A_n^{\rm MHV} &= \frac{\langle ij\rangle^4}{\langle 12 \rangle \langle 23 \rangle \ldots \langle n1 \rangle}
\sim \frac{(z_i-z_j)^4}{(z_1-z_2)(z_2-z_3) \ldots (z_n-z_1)} .
}
Here the collinear singularities are manifest in the form of $z_i - z_{i+1}$ poles in the denominator.  More generally, since projective spinors only carry angular information, they are useful for exposing the collinear behavior of expressions.

\subsection{Approach}

\label{sec:approach}

So far we have simply defined a convenient representation of 4D Minkowski space as Milne and Rindler regions foliated into warped 3D slices.  While at last  we appear poised to apply the $\AdS_3/\CFT_2$ dictionary, a naive ambiguity arises: $\Milne_4$ reduces to a {family} of $\AdS_3$ slices---to which should we apply the holographic correspondence?  After all, each value of $\tau$ corresponds to a distinct $\AdS_3$ geometry, each with a different curvature and position in $\Milne_4$.  Even stranger, the bulk dynamics of $\Mink_4$ will in general intersect all foliations of both $\Milne_4$ and $\Rind_4$.

The resolution to this puzzle is rather straightforward---and ubiquitous in more conventional applications of $\AdS/\CFT$. Perhaps most familiar is the case of spacetimes with factorizable geometry, $\AdS \times {\cal M}$, where ${\cal M}$ is a compact manifold.  In such circumstances, the appropriate course of action is to Kaluza-Klein (KK) reduce the degrees of freedom along the compact directions of  ${\cal M}$.  This generates a tower of KK modes in AdS to which the standard AdS/CFT dictionary  should then be applied. In a slightly more complicated scenario, the spacetime is a { warped} product of AdS and ${\cal M}$, where the AdS radius varies from point to point in ${\cal M}$. Here too, KK reduction to AdS---with some fiducial radius of curvature---can be performed, again resulting in a tower of KK modes.  

Something very similar occurs in our setup because $\Milne_4$ is simply a warped product of AdS$_3$ and $\R_{\tau}$, the real line parameterizing Milne time. Here ``KK reduction'' corresponds to a decomposition of fields in $\Milne_4$ into modes in Milne time $\tau$ which are in turn 
AdS$_3$ fields via separation of variables. Each mode is then interpreted as a separate particle residing in the dimensionally reduced $\AdS_3$.    However, unlike the usual KK scenario, where the spectrum of particles is discrete, the non-compactness of $\R_{\tau}$ induces a continuous ``spectrum'' of $\AdS_3$ modes.
    As we will see later, an effective ``compactification'' \cite{stromingerKacMoody} occurs when we consider the soft limit, which is the analog of projecting onto zero modes in the standard Kaluza-Klein procedure.  

In the subsequent sections we derive this mode decomposition for scalar and gauge theories in the Milne region.  We consider theories that exhibit classical Weyl invariance, permitting  $\Milne_4$ to be recast as a nicely factorized geometry, $\AdS_3 \times \R_\tau$, rather than a warped product. In this case the mode decomposition is especially simple because Milne energy is conserved.  Note, however, that this is merely a technical convenience that is not essential for our main results.  In particular, when we go on to consider the case of gravity, there will be no such Weyl invariance, but the reduction of $\Milne_4$ down to $\AdS_3$ modes is of course still possible.

Armed with a reduction of  $\Milne_4$ degrees of freedom down to $\AdS_3$, we then apply the $\AdS_3/\CFT_2$ dictionary to recast scattering amplitudes in the form of $\CFT_2$ correlators. We then show how the embedding formalism offers a trivial continuation of these results from $\Milne_4$ into $\Rind_4$ and thus all of $\Mink_4$.  Along the way, we will understand the 4D interpretation of familiar objects in the $\CFT_2$, including correlators, Ward identities, and current algebra.

\section{Gauge Theory}

\label{sec:gauge}

\subsection{Mode Expansion from $\Milne_4$ to $\AdS_3$}

\label{sec:gaugemodeexpansion}

As a simple warmup, consider the case of a massless interacting scalar field in Minkowski space.  For the sake of convenience, we focus on Weyl invariant theories, although as noted previously this is not a necessity.  The simplest Weyl invariant action of a scalar is 
\eq{
S &= \int_{\Milne_4} d^4Y \, \sqrt{-G} \left(-\frac{1}{2}  G^{IJ} \nabla_I \Phi \nabla_J \Phi   - \frac{1}{12} R \Phi^2   - \frac{\lambda}{24} \Phi^4\right),
\label{eq:scalar_action}
}
for now restricting to the contribution to the action from $\Milne_4$.  An identical analysis will apply to $\Rind_4$, and later we will discuss at length how to glue these regions together.

In \Eq{eq:scalar_action} the conformal coupling to the Ricci scalar has no dynamical effect in flat space because $R=0$.  Nevertheless, this interaction induces an improvement term in the stress tensor for the scalar that ensures Weyl invariance.    The Weyl transformation is given by 
\eq{
G_{IJ} &\rightarrow  \bar G_{IJ}= e^{-2\tau} G_{IJ},
\label{eq:conftransformation}
}
where the scalar transforms as 
\eq{
\Phi \rightarrow \bar{\Phi} = e^{\tau} \Phi.
}
 Due to the classical Weyl invariance of the theory, the metric decomposes into a factorizable $\AdS_3 \times \R_\tau$ geometry with the associated metric,
\eq{
d s^2_{\AdS_3 \times \R_\tau} = \bar{G}_{IJ} dY^I dY^J = -d \tau^2 + ds^2_{\AdS_3}, 
}
where $ds^2_{\AdS_3}$ is defined in \Eq{eq:hyper_coords}.  Since the action is Weyl invariant we obtain 
\eq{
S &= \int_{\AdS_3} d^3y\, \sqrt{-\bar G} \int d\tau \left(-\frac{1}{2}  \bar G^{IJ} \nabla_I \bar{\Phi} \nabla_J \bar{\Phi} - \frac{1}{12} \bar R \bar{\Phi}^2 - \frac{\lambda}{24} 
\bar{\Phi}^4    \right),
}
where  $\bar R = -6$ is the curvature of the $\bar G_{IJ}$ metric. 

Given the factorizable geometry, it is natural to define a ``Milne energy'',
\eq{
\omega = i {\partial_\tau},
\label{eq:Milne_energy}
}
which is by construction a Casimir invariant under the $\AdS_3$ isometries, or in the language of the dual $\CFT_2$, the global conformal group $SL(2, \mathbb{C})$. This $SL(2, \mathbb{C})$ is also the 4D Lorentz group acting on the $\Milne_4$ embedding space of $\AdS_3$. By contrast, the usual Minkowski energy,
\eq{
E = i {\partial_0},
}
is of course not Lorentz invariant and thus not $SL(2, \mathbb{C})$ invariant, and so is less useful in identifying the underlying $\CFT_2$ structure. Again, we emphasize here that the Weyl invariance of the scalar theory is an algebraic convenience that is not crucial for any of our final conclusions.  When Weyl invariance is broken, then the Milne energy simply is not conserved.  

We can now expand the scalar into harmonics in Milne time,
\eq{
\phi(\omega) &= \int d\tau \, e^{i\omega \tau} \bar{\Phi}(\tau),
}
where $\phi(\omega)$ are scalar fields in $\AdS_3$, analogous to the tower of KK modes that arise in conventional compactifications.  In terms of these fields, the linearized action becomes
\eq{
S_0 &= \int_{\AdS_3} d^3y \, \sqrt{g}  \int d \omega  \left( -\frac{1}{2} g^{ij} \nabla_i  \phi(-\omega) \nabla_j \phi(\omega)   + \frac{1}{2}  (1 + \omega^2) 
 \phi(-\omega)   \phi(\omega) \right),
}
so a massless scalar field in $\Milne_4$ decomposes into a tower of $\AdS_3$ scalars with
\eq{  
m_{\phi}^2(\omega) = -(1+\omega^2).  \label{eq:scalarmass}
}
Curiously, the mass violates the 3D Breitenlohner-Freedman bound \cite{bfBound1,bfBound2} and is thus formally tachyonic in $\AdS_3$.  In fact,  as the Milne energy grows, the mass becomes more tachyonic simply because we have mode expanded in a time-like direction. While such pathologies ordinarily imply an unbounded from below Hamiltonian, one should realize here that the $\AdS_3$ theory is Euclidean and the true time direction actually lies outside the warped geometry.

Next, let us proceed to the case of 4D gauge theory.  We consider the YM action, 
\eq{
S &= -\frac{1}{2g_{\rm YM}^2}  \int_{\Milne_4} d^4Y \, \sqrt{-G} \, {\rm tr}\left( G^{IJ} G^{KL}F_{IK} F_{JL}  \right),
}
again focusing on contributions from the Milne region.  Here $F_{IJ}$ is the Lie algebra-valued non-abelian gauge field strength.
Under a Weyl transformation, the metric transforms according to \Eq{eq:conftransformation}, while the gauge field is left invariant,
\eq{
 A_I \rightarrow A_I.
}
Due to the classical Weyl invariance of 4D YM theory, this transformation leaves the action unchanged, so\eq{
S &= -\frac{1}{2g_{\rm YM}^2}  \int_{\AdS_3} d^3y\, \sqrt{-\bar G} \int d\tau \, {\rm tr} \left(  \bar G^{IJ} \bar G^{KL}  F_{IK}  F_{JL}  \right).
}
As before, the Weyl invariance of the action is a convenience whose main purpose is to simplify some of the algebra.

Decomposing the gauge field as $A_I = (A_\tau , A_i)$ and going to Milne temporal gauge, $A_\tau = 0$, we rewrite the linearized action as
\begin{equation}
S_0 = \frac{1}{g_{\rm YM}^2} \int_{\AdS_3} d^3y\,  \sqrt{g}\int d \omega  \, {\rm tr} \left( -\frac{1}{2} g^{ij} g^{kl}  f_{ik}(-\omega) f_{jl}(\omega) + 
\omega^2 \gamma^{ij} a_i(-\omega)a_j(\omega)  \right), \label{eq:Milnecompact}
\end{equation}
where $f_{ij} = \partial_i a_j - \partial_j a_i$ is the linearized field strength associated with the Milne modes,
\eq{
a_i(\omega) = \int d\tau \, e^{i\omega \tau} A_i(\tau).
}
From \Eq{eq:Milnecompact} we see that the $a_i(\omega)$ are Proca vector fields in $\AdS_3$ with mass
 \eq{
 m_{a}^2 (\omega)= - \omega^2. \label{eq:vectormass}
 }
The $\AdS_3$ fields are formally tachyonic since we have mode expanded in the time-like Milne direction.  In summary, we find that a massless vector in $\Milne_4$ decomposes into a continuous tower of massive Proca vector fields in $\AdS_3$.

\subsection{Scaling Dimensions from $\AdS_3/\CFT_2$}

\label{sec:scalingdimensions}

According to the standard holographic dictionary, each field in $\AdS_3$ is dual to a $\CFT_2$ primary operator with 
scaling dimension 
$\Delta$ dictated by the corresponding $\AdS_3$ mass.  From \Eq{eq:scalarmass} and \Eq{eq:vectormass}, we deduce that the scaling dimensions for scalar and vector primaries satisfy $\Delta_\phi(\Delta_\phi-2) = m^2_{\phi}(\omega)=-(1+\omega)^2$ and $(\Delta_a-1)^2 = m^2_{a}(\omega)=-\omega^2$.  Both equations imply the following relationship between the scaling dimension and the Milne energy,
\begin{equation}
\Delta(\omega) = 1 \pm i \omega. 
 \label{eq:Deltadef}
\end{equation}
Since unitary CFTs and their Wick-rotated Euclidean versions have real scaling dimensions, the CFT encountered here is formally non-unitary.  This is true despite the manifest unitarity of the underlying 4D dynamics.

\subsection{Witten Diagrams in $\AdS_3$}

\label{sec:gaugewittendiagrams}

With the mode decomposition just discussed, it is a tedious but straightforward exercise to derive an explicit action for the tower of $\AdS_3$ modes descended from $\Milne_4$.  From this action we can then compute Witten diagrams in $\AdS_3$.  By the $\AdS_3/\CFT_2$ dictionary, these Witten diagrams are equivalent to correlators of a certain $\CFT_2$.  As we will argue here and in subsequent sections, these Witten diagrams and correlators are also equal to scattering amplitudes in $\Mink_4$.  

A priori, such a correspondence is quite natural.  First of all, tree-level Witten diagrams and scattering amplitudes both describe a classical minimization problem---{\it i.e.}~finding the saddle point of the action subject to a particular set of boundary conditions.  Second, the $\CFT_2$ resides on the $\dAdS_3$ boundary, which at $\tau\rightarrow +\infty$ houses massless asymptotic in and out states.  

In any case, we will derive an explicit mapping between the basic components of Witten diagrams and scattering amplitudes.  The former are comprised of interaction vertices, bulk-bulk propagators, and bulk-boundary propagators, while the latter are comprised of interaction vertices, internal propagators, and a prescription for LSZ reduction.    Let us analyze each of these elements in turn.

\subsubsection{Interaction Vertices}

To compute the interaction vertices of the $\AdS_3$ theory we simply express the interactions in $\Milne_4$ in terms of the mode decomposition into massive $\AdS_3$ fields.  For example, the quartic self-interaction of the scalar field becomes
\eq{
S_{\rm int} &=  - \frac{\lambda}{24} \int_{\AdS_3} d^3y\,  \sqrt{g}\int d \omega_1 d \omega_2 d \omega_3 d \omega_4 \, \phi(\omega_1)\phi(\omega_2)\phi(\omega_3)\phi(\omega_4) \delta(\omega_1+\omega_2+\omega_3+\omega_4)  ,
}
so interactions in the bulk of $\Milne_4$ translate into interactions among massive scalars in $\AdS_3$.  Due to the Weyl invariance of the original scalar theory, these interactions conserve Milne energy.

It is then clear that the interaction vertices of 3D Witten diagrams are equivalent to those of 4D flat space Feynman diagrams modulo a choice of coordinates---that is, Milne versus Minkowski coordinates, respectively.   While these Witten diagram interactions typically involve complicated interactions among many $\AdS_3$ fields, this is just a repackaging of standard Feynman vertices.

\subsubsection{Bulk-Bulk Propagator}

In this section we show that the bulk-bulk propagators of Milne harmonics in $\AdS_3$ are simply a repackaging of Feynman propagators in $\Mink_4$.  To simplify our discussion, let us again revisit the case of the massless scalar field, although a parallel discussion holds for gauge theory but with the extra complication of gauge fixing.  

Consider the Feynman propagator for a massless scalar field in flat space, 
\begin{equation}
G(\tau, y, \tau', y')_{\Mink_4} = \frac{ i }{ \Box_{\Mink_4}} = e^{-\tau'} \frac{i}{\Box_{\AdS_3} + 1 -\partial_{\tau}^2} e^{- \tau}, 
\end{equation}
where $(\tau,y)$ and $(\tau',y')$ are points in the Milne region.  Here we have defined 
\eq{
\Box_{\Mink_4} = \nabla_I \nabla^I \qquad \textrm{and} \qquad
\Box_{\AdS_3} = \nabla_i \nabla^i,
}
to be the d'Lambertian in $\Mink_4$ and the Laplacian in $\AdS_3$, respectively.   This expression is manifestly of the form of the $\AdS_3$ propagator with $e^{-\tau}$ factors inserted to account for the non-trivial Weyl weight of the scalar field.  Indeed, by applying the Weyl transformation and decomposing into Milne modes, we obtain the $\AdS_3$ propagator for a scalar,
\begin{equation}
G(\omega, y,y')_{\AdS_3} =
\frac{i}{\Box_{\AdS_3} + 1 + \omega^2},
\end{equation} 
which automatically satisfies the wave equation for a scalar in $\AdS_3$,
\eq{
(\nabla_i \nabla^i+ 1+  \omega^2)G(\omega, y,y')_{\AdS_3} = i\delta^3(y,y').
}
Hence, the Feynman propagator is a particular convolution over a tower of $\AdS_3$ propagators.

Of course, the above statements are purely formal until the differential operator inverses are properly defined by an $i \epsilon$ prescription. 
The Minkowski propagator takes the usual $i \epsilon$ prescription, 
\begin{equation}
G(\tau,y, \tau', y')_{\Mink_4} = \frac{ i }{ \Box_{\Mink_4} + i \varepsilon},
\end{equation}
which selects the Minkowski vacuum as the ground state of the theory. This is, however, not the natural vacuum of the Weyl-transformed geometry, $\AdS_3 \times \R_{\tau}$, which is instead the {\it conformal vacuum} corresponding to the ground state with respect to the Milne Hamiltonian, {\it i.e.}~$\tau$ translations. In order to match the propagator of the Minkowski vacuum we must choose the {\it thermal propagator} in $\AdS_3 \times \R_{\tau}$ \cite{birrellDavies}.  Thermality arises from the entanglement between the Milne and Rindler regions of Minkowski spacetime across the Rindler horizon $x^2 = 0$. With this prescription, Feynman propagators in $\Mink_4$ can be matched directly to bulk-bulk propagators in $\AdS_3$.   Note that thermality does not break the $SL(2,\mathbb{C})$ Lorentz symmetries, since these act only on the $\AdS_3$ coordinates and not the Milne time or energy.

A similar story holds for gauge fields.  Going to Milne temporal gauge, the $\Mink_4$ gauge propagator can be expressed as a convolution over massive $\AdS_3$ Proca propagators. These propagators satisfy the Proca wave equation sourced by a delta function,  
\begin{equation}
(\nabla_k \nabla^k\delta_i^{\;j} - \nabla_i \nabla^j +  \omega^2 \delta_i^{\; j})G_{jl}(\omega, y,y')_{\AdS_3} = i\delta_{il} \delta^3(y,y'),
\label{eq:procaEq}
\end{equation}
where we have Fourier transformed to Milne harmonics.

\subsubsection{Bulk-Boundary Propagator}

We have now verified that the bulk interaction vertices and bulk-bulk propagators of Witten diagrams in $\AdS_3$ are simply Feynman diagrammatic elements in the $\Milne_4$ embedding space. The final step in matching Witten diagrams to scattering amplitudes is to match their respective boundary conditions.  For Witten diagrams, the external lines are  $\AdS_3$ bulk-boundary propagators.  For scattering amplitudes, the external lines are fixed by LSZ reduction to be solutions of the Mink$_4$ free particle equations of motion---taken usually to be plane waves.  Here we derive a concrete relationship between the bulk-boundary propagators and LSZ reduction.  

To begin, let us compute the bulk-boundary propagator for primary fields of scaling dimension $\Delta$. At this point it will be convenient to employ the elegant embedding formalism of \cite{penedones}, which derived formulas for the bulk-boundary propagator in terms of a flat embedding space of one higher dimension.  
Ordinarily, AdS is considered physical while the flat embedding space is an abstraction devised to simplify the bookkeeping of curved spacetime. Here the scenario is completely reversed: flat space is physical while AdS is the abstraction introduced in order to recast flat space dynamics into the language of CFT.

In the embedding formalism \cite{penedones}, the bulk-boundary propagator for a scalar primary is
\eq{
K^{\Delta} = \frac{1}{(k x)^{\Delta}}.
\label{eq:scalarBB}
}
Since we have lifted from $\AdS_3$ to $\Mink_4$, the right-hand side actually depends on 4D quantities.  Specifically, the four-vector $x^\mu$ labels a point in $\Mink_4$ while the four-vector $k^\mu$ labels a point $(z,\bar z)$ on the boundary of $\AdS_3$ according to \Eq{eq:nq_def}.

Already, we see an elegant subtlety that arises in the embedding formalism: each point in $\AdS_3$ is recast as a point in $\Mink_4$ with the implicit constraint $x^2 = - 1$.  In Milne coordinates, this corresponds to the constraint $\tau = 0$. We can, however, ``lift'' the bulk-boundary propagators from $\AdS_3$ to $\Mink_4$ by simply dropping this constraint, yielding a bulk-boundary propagator with an additional $\tau$ dependent factor, $e^{- \tau \Delta}$.  Combined with an extra factor of $e^\tau$ for the Weyl weight of a scalar field, this generates a net phase $e^{-\tau (\Delta-1)} = e^{\mp i \omega \tau }$ from the definition of $\Delta$ in \Eq{eq:Deltadef}.  We immediately recognize this as the phase factor that accompanies the Fourier transform between $\tau$ dependent fields in $\AdS_3 \times  \R_{\tau}$ and $\omega$ dependent Milne harmonics. That is, the lifted propagators can be used to compute the boundary correlators of modes in $\AdS_3$ in terms of boundary correlators of 4D states in $\AdS_3 \times  \R_{\tau}$. The fact that the bulk-boundary propagators satisfy the free particle equations of motion in $\AdS_3$  translates to the fact that the Weyl-transformed lifted propagators satisfy the free particle equations of motion in $\AdS_3 \times  \R_{\tau}$ via separation of variables.  In turn, this implies that the embedding formalism bulk-boundary propagator in \Eq{eq:scalarBB} satisfies the equations of motion in $\Mink_4$. This fact is straightforwardly checked by direct computation.

Next, consider the bulk-boundary propagator for a vector primary, $K^{\Delta}_i$.  This object is fundamentally a bi-vector since it characterizes propagation of a vector disturbance from the $\dAdS_3$ boundary into the bulk of $\AdS_3$.  While the 3D bulk vector index is manifest, the 2D boundary vector index is suppressed---implicitly taken here to be either the $z$ or $\bar z$ component.  As for the scalar, we can lift the $\AdS_3$ bulk-boundary propagator to $K^{\Delta}_I = (K^{\Delta}_\tau , K^{\Delta}_i)$ where we assume Milne temporal gauge to set $K^{\Delta}_\tau=0$.  Going to Minkowski coordinates, we obtain
\eq{
 K_{\mu}^{\Delta}  = \frac{\partial y^I}{\partial x^\mu} K^{\Delta}_I  = \frac{1}{(k x)^{\Delta}} \left(\epsilon_\mu- \frac{\epsilon x}{kx} k_\mu \right),
\label{eq:KDeltadef}
}
where we have chosen the $z$ component of the boundary vector.
Here the dependence on boundary coordinates $(z,\bar z)$ enters through $k$ and $\epsilon$ according to \Eq{eq:nq_def} and \Eq{eq:eebar_def}.  Had we instead chosen the $\bar z$ component of the boundary vector, we would have obtained the same expression as \Eq{eq:KDeltadef} except with $\bar\epsilon$ instead of $\epsilon$.

\subsection{Continuation from $\Milne_4$ to $\Mink_4$}

\label{sec:continuation}

Until now, the ingredients of our discussion---interaction vertices, bulk-bulk propagators, and bulk-boundary propagators---have all been restricted to Milne region time-like separated from the origin.  However, it is clear that scattering processes in general will also  involve the Rindler region space-like  separated from the origin.  
 As we will see, this is not a  problem because the Milne diagrammatic components---written in terms of flat space coordinates via the embedding formalism---can be trivially continued to the Rindler region and thus all of Minkowski spacetime.
 
To be concrete, recall the foliation of the Rindler region in \Eq{eq:Rindler_x2} and \Eq{eq:Rindler_metric}.  
Each slice of constant $\rho$ defines a Lorentzian $\dS_3$ spacetime.
In $\Rind_4$, boundary correlators correspond to Witten diagrams of $\dS_3$ fields descended from a mode decomposition with respect to the Rindler {\it momentum},  
$\omega = i \partial_{\rho}$.  Moreover, the lifted bulk-boundary propagators in $\Rind_4$ are given precisely by \Eq{eq:scalarBB} and \Eq{eq:KDeltadef}, except continued to the full $\Mink_4$ region for any value of $x^2$.  So the embedding formalism gives a perfect prescription for continuation from Milne to Rindler.   One can also think of this as a simple analytic continuation of the original $\AdS_3$ theory into $\dS_3$, which shares the same $SL(2, \mathbb{C})$ Lorentz isometries.

This result implies that $\Mink_4$ scattering amplitudes---properly LSZ-reduced on bulk-boundary propagators on both the Milne and Rindler regions---are equal to a 3D Witten diagrams for Milne and Rindler harmonics which splice together boundary correlators in $\AdS_3$ and $\dS_3$.  Using these continued Witten diagrams, we can then define a set of $\CFT_2$ correlators dual to scattering amplitudes through a hybrid of the $\AdS_3/\CFT_2$ and $\dS_3/\CFT_2$ \cite{stromingerdSCFT,wittenQGindS,maldacenaNonGaussianities,stromingerHigherSpin} correspondences. Note that the smooth match between correspondences, given the Euclidean signature of $\AdS_3$ and the Lorentzian signature of $\dS_3$.

 As a consequence, our proposed correspondence between $\Mink_4$ and $\CFT_2$ is subtle. While the Minkowski theory is unitary, the $\CFT_2$ is not unitary in any familiar sense---a fact which is evident from the appearance of complex scaling dimensions in \Eq{eq:Deltadef}.   This is not a contradiction, since unlike the usual $\AdS_3/\CFT_2$ correspondence, the time direction and unitary evolution are emergent, as in the spirit of $\dS_3/\CFT_2$.  The question of how flat space unitarity is encoded within  a non-unitary CFT obviously deserves further study.

 \subsection{$\Mink_4$ Scattering Amplitudes as $\CFT_2$ Correlators}
 
 \label{sec:MAdSCFT}
 
 Assembling the various diagrammatic ingredients, we see that Witten diagrams for the $\textrm{(A)dS}_3$ fields descended from the mode decomposition of $\Mink_4$ are equal to 4D scattering amplitudes---albeit with a modified prescription for LSZ reduction in which the usual external wavepackets of fixed momentum are replaced with the lifted bulk-boundary propagators of \Eq{eq:scalarBB} and \Eq{eq:KDeltadef}. These alternative ``wavepackets'' may seem unfamiliar, but crucially, they can 
 be expressed as superpositions of on-shell plane waves.
 
 For the scalar field this is straightforward, since the bulk-boundary propagator in \Eq{eq:scalarBB} can be expressed as a Mellin transform of plane waves \cite{deBoerSolodukhin}, 
 \eq{
K^{\Delta} = \frac{1}{(kx+i\varepsilon)^\Delta} = \frac{i^{-\Delta}}{\Gamma(\Delta)} \int_0^\infty ds \, s^{\Delta-1} e^{iskx}e^{-\varepsilon s},
\label{eq:Kintegral}
}
where $\varepsilon$ is an infinitesimal regulator.  Here the right-hand side is manifestly a superposition of on-shell plane waves, $e^{iskx}$, since $k^2=0$.  

Something similar happens for the gauge field since
 \eq{
K^{\Delta}_\mu = \left(\epsilon_\mu + \frac{k_\mu \partial_z}{\Delta} \right) \frac{1}{(kx)^{\Delta}}.
\label{eq:Kmuintegral}
}
Using the simple observation that $k_\mu \partial_z (\cdot) = \partial_z(k_\mu  \cdot)  - \epsilon_\mu (\cdot)$, we see that \Eq{eq:Kintegral} and \Eq{eq:Kmuintegral}  imply that $K_\mu^{\Delta}$ is a superposition of on-shell plane waves, $\epsilon_{\mu} e^{iskx}$, up to a superposition of pure gauge transformations, 
 $k_{\mu} e^{iskx}$.

\begin{figure}[t]
\begin{center}
\includegraphics[width=.95\textwidth]{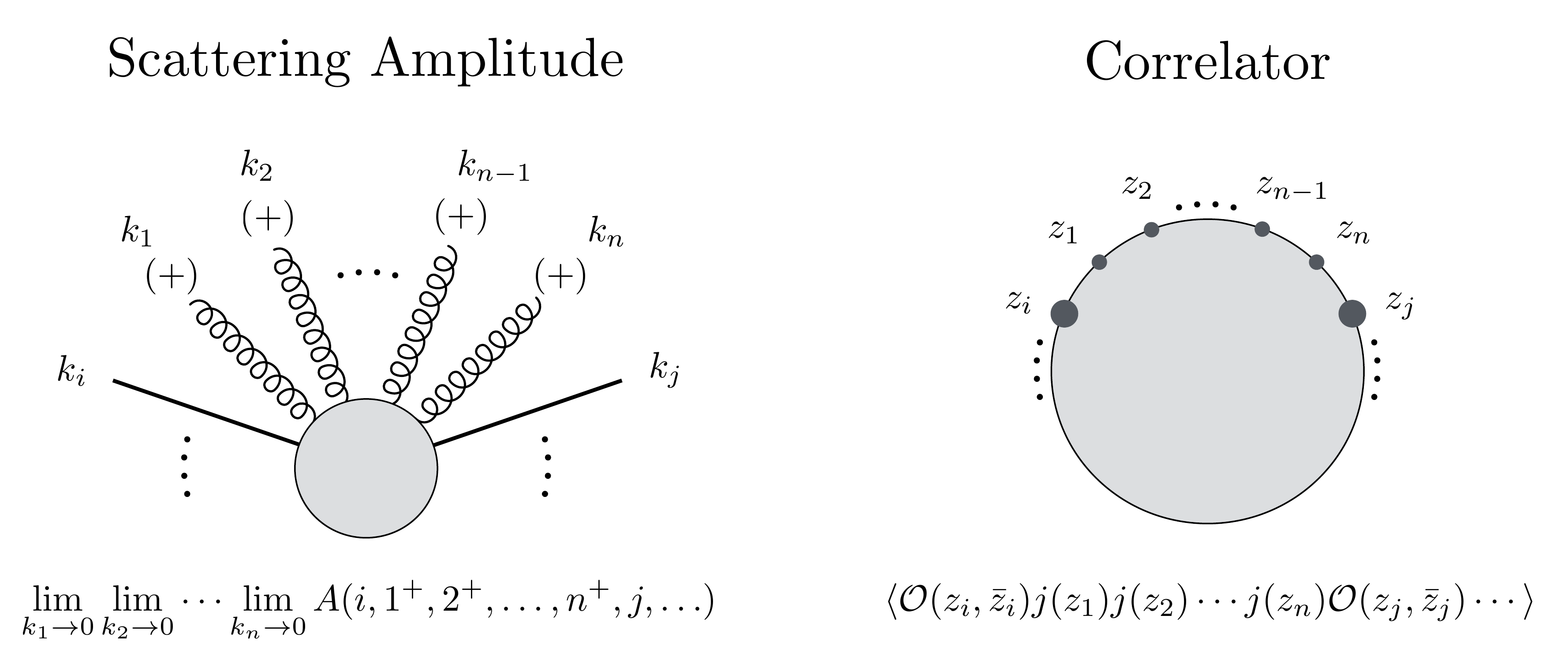}
\end{center}
\vspace*{-0.5cm}
\caption{Equivalence of 4D scattering amplitudes and 2D correlators for the special case of multiple soft boson gauge emission and multiple conserved current insertion.     }
 \label{fig:correspondence}
\end{figure}

In this way, we have shown that every Witten diagram can be written as a superposition of on-shell scattering amplitudes in $\Mink_4$, or equivalently as a single scattering amplitude with a modified LSZ-reduction to certain bulk-boundary wavepackets.   By the (A)dS/CFT dictionary, this implies that the latter are equivalent to Euclidean correlators of a $\CFT_2$ on the $\partial \textrm{(A)dS}_3$ boundaries, which together form the entirety of past and future null infinity.  Concretely, this implies the equivalence of correlators and scattering amplitudes,
\begin{equation}
  \langle \mathcal{O}^{\Delta_1}(z_1, \bar{z}_1) \cdots \mathcal{O}^{\Delta_n}(z_n, \bar{z}_n) \rangle = A(K^{\Delta_1}(z_1,\bar z_1),\ldots, K^{\Delta_n}(z_n,\bar z_n))  = \langle \textrm{out} | \textrm{in}\rangle,
\label{eq:mainresult}
\end{equation}
where here we have restricted to scalar operators for simplicity, but the obvious generalization to higher spin applies.
In \Eq{eq:mainresult} the quantity $A$ denotes a scattering amplitude with a modified LSZ-reduction replacing the usual plane waves with the lifted bulk-boundary propagators $K^{\Delta_i}(z_i,\bar z_i)$ corresponding to the boundary operators ${\cal O}^{\Delta_i}(z_i,\bar z_i)$.  The associated scaling dimension of each operator is $\Delta_i = 1+i\omega_i$, and if the bulk theory is conformally invariant in 4D, for example as in massless gauge theory at tree level, then $\sum_{i=1}^n \omega_i=0$.  The boundary operators are naturally divided into two types, $\mathcal{O}_\text{in}$ and $\mathcal{O}_\text{out}$, depending on sign of the Minkowski energy $E > 0$ or $E < 0$,  corresponding to scattering states that are incoming or outgoing, respectively. This equivalence of correlators and scattering amplitudes is depicted in \Fig{fig:correspondence}.

\subsection{Conserved Currents of $\CFT_2$}

\label{sec:conservedcurrents}

In \Eq{eq:mainresult}, we derived an explicit holographic correspondence between scattering amplitudes in $\Mink_4$ and correlators of a certain $\CFT_2$.  For gauge fields, the associated massive 
AdS$_3$ modes are dual to {\it non-conserved} currents in the $\CFT_2$ while the massless $\AdS_3$ modes are dual to {\it conserved} currents in the $\CFT_2$.  Since the mass of an $\AdS_3$ vector is proportional to its Milne energy by \Eq{eq:vectormass}, we can study the massless case by taking the limit of vanishing Milne energy $\omega=0$, {\it i.e.}~the Milne soft limit.   For the dual vector primary operator, this corresponds to $\Delta = 1$, so the correlator reduces to the Ward identity for current conservation in the $\CFT_2$.

To start, consider the bulk-boundary propagator for a massless $\AdS_3$ vector, 
\eq{
K_\mu =\frac{x^\rho f_{\rho \mu}  }{(kx)^{2}} ,
\label{eq:K_gauge}
}
obtained by setting $\Delta = 1$ in  \Eq{eq:KDeltadef}.  Here we have defined linearized field strengths constructed from boundary data,
\eq{
f_{\mu\nu}  = k_\mu \epsilon_\nu - k_\nu \epsilon_\mu ,\qquad \textrm{and} \qquad \bar f_{\mu\nu}  = k_\mu \bar\epsilon_\nu - k_\nu \bar\epsilon_\mu.
\label{eq:fdefs}
}
Note that $x^\mu K_\mu = K_\tau = 0$ since we have chosen Milne temporal gauge.
Remarkably,  $K_\mu$ is actually a total derivative with respect to  $\Mink_4$ coordinates, 
\eq{
K_\mu =  \partial_\mu \xi   \qquad \textrm{where} \qquad \xi =  \frac{\epsilon x}{kx}.
\label{eq:pure_gauge_gauge}
}
This fact dovetails beautifully with the results of \cite{stromingerFirstPaper,stromingerQED,stromingerMassiveQED,stromingerMagnetic}, which argued that there is physical significance to large gauge transformations that do not vanish at the boundary of $\Mink_4$.    
As we will see, concrete calculations are vastly simplified using the pure gauge form of $K_\mu$.

\subsubsection{$\Mink_4$ Soft Theorems as $\CFT_2$ Ward Identities}

Let us start with the simplest case of abelian gauge theory with arbitrary charged matter.  
We showed earlier that a $\Mink_4$ scattering amplitude with a Milne soft gauge boson can be expressed as a Witten diagram for a massless $\AdS_3$ vector field,
\eq{
\langle j(z) {\cal O}(z_1,\bar z_1)\cdots {\cal O}(z_n,\bar z_n) \rangle = \int d^4 x \, K_\mu(x) W^\mu(x).
\label{eq:Smatrixgauge}
}
Here the left-hand side is a correlator involving the $\Delta=1$ conserved current of the $\CFT_2$ and $K_\mu$ is the bulk-boundary propagator for the massless vector in $\AdS_3$. The function $W^\mu$ represents the remaining contributions to the Witten diagram from bulk interactions,
\eq{
W^\mu(x) = \langle \textrm{out} |J^\mu(x) |\textrm{in} \rangle,
\label{eq:Wdef}
}
where $J^\mu$ is the gauge current operator of 4D Minkowski spacetime inserted between scattering states.   Here the in and out states are defined according to the modified prescription for LSZ reduction shown in \Eq{eq:mainresult}.

Inserting the pure gauge form of $K_\mu$ in \Eq{eq:pure_gauge_gauge} and integrating by parts, we obtain
\eq{
\langle j(z) {\cal O}(z_1,\bar z_1)\cdots {\cal O}(z_n,\bar z_n)  \rangle &=  \int d^4 x\, \partial_\mu\xi(x)  \langle \textrm{out} |J^\mu(x) |\textrm{in} \rangle \nonumber  \\
&=  -\int d^4 x\, \xi(x) \partial_\mu  \langle \textrm{out} |  J^\mu(x) |\textrm{in} \rangle.
}
By dropping total derivatives, we have implicitly assumed that $W^\mu$ describes a charge configuration that vanishes on the boundary.  Naively, this stipulation is inconsistent if the bulk process involves charged external particles that propagate to the asymptotic boundary.  However, this need not be a contradiction, provided $W^\mu$ is sourced by insertions of charged particles near but not quite on the boundary.   Conservation of charge is effectively violated wherever the external particles are inserted, so
\eq{
 \partial_\mu\langle \textrm{out} | J^\mu(x) |\textrm{in} \rangle = -\sum_{i=1}^n q_i \delta^4(x-x_i)\langle \textrm{out} | \textrm{in} \rangle.
}
Here $i$ runs over all the particles in the scattering process, $q_i$ are their charges, and $x_i$ are their insertion points near the $\dAdS_3$ boundary.  Crucially, we recall from \Eq{eq:x_expand} that massless particles near the $\dAdS_3$ boundary are located at positions $x_i$ that are aligned with their associated on-shell momenta, $k_i$.  This is simply the statement that the positions of asymptotic states on the celestial sphere point in the same directions as their momenta.  In any case, the upshot is that as $\rho_i \rightarrow 0$, we can substitute $x_i \sim k_i$.  

Plugging in \Eq{eq:pure_gauge_gauge} and \Eq{eq:mainresult}, and replacing $x_i \sim k_i$, we can trivially integrate the delta function to obtain 
\eq{
\langle j(z) {\cal O}(z_1,\bar z_1)\cdots {\cal O}(z_n,\bar z_n)\rangle =  \sum_{i=1}^n  q_i \left(\frac{\epsilon k_i}{k k_i}\right)\langle {\cal O}(z_1,\bar z_1)\cdots {\cal O}(z_n,\bar z_n)  \rangle,
\label{eq:Weinbergsoft}
}
which is exactly the Weinberg soft factor for soft gauge boson emission \cite{weinberg}.  Here it was important that we identified $x_i \sim k_i$ so that the resulting Weinberg soft factor depends on the on-shell momenta, $k_i$.  Later on, we will occasionally find it useful to switch back and forth between the position and momentum basis for the hard particles.

At the same time, this expression simplifies further because
\eq{
\frac{\epsilon k_i}{k k_i} 
=\frac{1}{z-z_i},
\label{eq:exkx}
}
yielding the Ward identity for a 2D conserved current,
\eq{
\langle j(z) {\cal O}(z_1,\bar z_1)\cdots {\cal O}(z_n,\bar z_n)  \rangle =\sum_{i=1}^n \frac{q_i}{z-z_i}\langle  {\cal O}(z_1,\bar z_1)\cdots {\cal O}(z_n,\bar z_n)  \rangle.
\label{eq:softgauge}
 }
So \Eq{eq:Weinbergsoft} is simultaneously the soft theorem in $\Mink_4$, the Witten diagram for a massless vector in $\AdS_3$, and the Ward identity for a conserved current in the $\CFT_2$.  From this result we deduce that an insertion of the CFT conserved current is dual to a soft gauge boson emission. 

The above analysis for abelian gauge theory  is straightforwardly extended to the non-abelian case.  The equation for approximate current conservation instead becomes
\eq{
 \partial_\mu\langle \textrm{out} | J^{a\mu}(x) |\textrm{in} \rangle = -\sum_{i=1}^n \delta^4(x-x_i)\langle \textrm{out} |T^a| \textrm{in} \rangle
}
so again plugging in $x_i \sim k_i$, we generalize  \Eq{eq:softgauge} to
\eq{
\langle j(z)^a {\cal O}^{b_1}(z_1,\bar z_1)\cdots {\cal O}^{b_n}(z_n,\bar z_n)  \rangle &=  \sum_{i=1}^n  f^{a b_i c_i} \left(\frac{\epsilon k_i}{k k_i}\right)\langle  {\cal O}^{b_1}(z_1,\bar z_1)\cdots {\cal O}^{c_i}(z_i,\bar z_i) \cdots {\cal O}^{b_n}(z_n,\bar z_n) \rangle \nonumber \\
&=\sum_{i=1}^n \frac{f^{ab_i c_i }}{z-z_i} \langle  {\cal O}^{b_1}(z_1,\bar z_1)\cdots {\cal O}^{c_i}(z_i,\bar z_i) \cdots {\cal O}^{b_n}(z_n,\bar z_n) \rangle ,
\label{eq:softgaugeNA}
}
which is the $\Mink_4$ soft theorem and the $\CFT_2$ Ward identity for non-abelian gauge theory.

The duality between soft gauge bosons and holomorphic currents has direct implications for scattering amplitudes.  For example, consider the correlator for a sequence of holomorphic currents wedged between two operator insertions,
\eq{
\langle {\cal O}^{a_i}(z_i,\bar z_i) j(z_1)^{a_1}  \cdots j(z_{n})^{a_{n}}{\cal O}^{a_j}(z_j,\bar z_j) \rangle.
}
Current conservation requires that this object be purely a holomorphic in the variables  $z_i$.  However, this expression can also be computed by sequential soft limits of an amplitude with two hard particles, yielding
\eq{
\frac{1}{(z_i-z_1)(z_1-z_2) \cdots (z_{n-1}-z_n)(z_n-z_j)},
\label{eq:multisoft}
}
which is the color-stripped amplitude for multiple soft emission.
To obtain this formula for the multiple leading soft limit it was important that sequential soft limits of single helicity gauge bosons commute when applied to color-stripped amplitudes.  The resemblance of \Eq{eq:multisoft} to the denominator of the Park-Taylor formula is not an accident: this form is required so that the only poles of the amplitude are collinear singularities. 

\subsubsection{Equivalence of $\Milne_4$ and $\Mink_4$ Soft Limits}

We have shown that the Ward identities of for 2D conserved currents are the same as the Weinberg soft theorems for 4D gauge theory \cite{weinberg}.  However, an astute reader will realize that the Weinberg soft theorems correspond to the limit of small Minkowski energy, $E=i\partial_0$ while our construction has centered on the Milne energy, $\omega=i\partial_\tau$ since it is an $SL(2,\mathbb{C})$ Lorentz invariant quantity.  Naively this is discrepant, but as we will now show, the Milne and Minkowski soft limits, $E\rightarrow 0$ and $\omega\rightarrow 0$, are one and the same.

To see why, we compute a correlator for a non-conserved current $j^{\Delta}(z)$ and take the limit towards $\Delta \rightarrow 1$ or equivalently, the Milne soft limit $\omega\rightarrow 0$.   The correlator to start is
\eq{
\langle j^{\Delta}(z)  {\cal O}(z_1,\bar z_1)\cdots {\cal O}(z_n,\bar z_n)\rangle = \int d^4 x \, K^{\Delta}_\mu(x) W^\mu(x).
\label{eq:jO_omega}
}
Here  $W_\mu$ is defined as in \Eq{eq:Wdef} and for $K^{\Delta}_\mu$ we plug in \Eq{eq:Kintegral} and \Eq{eq:Kmuintegral} to obtain
\eq{
\langle j^{\Delta}(z) {\cal O}(z_1,\bar z_1)\cdots {\cal O}(z_n,\bar z_n) \rangle  =  \frac{i^{-\Delta}}{\Gamma(\Delta)}\left(\epsilon_\mu + \frac{k_\mu \partial_z}{\Delta} \right) \int_0^\infty ds\, s^{\Delta-1} \langle \textrm{out} |\tilde J^\mu(sk) |\textrm{in} \rangle,
\label{eq:jO_simpTEMP}
}
where $\tilde J^\mu$ is the Fourier transform of $J^\mu$.  At this point we recognize $\tilde J^\mu$ as a Feynman diagram with an injection of momentum $sk$.   Notice that the integration variable $s$ has taken the role of the Minkowski energy of the inserted momentum.   The 4D Ward identity for on-shell gauge theory amplitudes is
\eq{
\langle \textrm{out} |k_\mu \tilde J^\mu(sk) |\textrm{in} \rangle=0,
}
whenever $\tilde J^\mu$ is evaluated at on-shell kinematics.  Again using $k_\mu \partial_z (\cdot) = \partial_z(k_\mu  \cdot)  - \epsilon_\mu (\cdot)$,   we are then permitted to reshuffle derivatives in  \Eq{eq:jO_simpTEMP}, where the first term on the right-hand side of this substitution vanishes by the Ward identity.  Doing so, we arrive at our final expression for the correlator,
\eq{
\langle j^{\Delta}(z)  {\cal O}(z_1,\bar z_1)\cdots {\cal O}(z_n,\bar z_n)\rangle  =  \frac{i^{-\Delta}(\Delta-1)}{\Gamma(\Delta+1)}  \int_0^\infty ds\, s^{\Delta-1} \langle \textrm{out} |\epsilon_\mu \tilde J^\mu(sk) |\textrm{in} \rangle.
\label{eq:jO_simp}
}
Since $\tilde J^\mu$ is evaluated at the on-shell momentum $sk$ and dotted into the on-shell polarization $\epsilon$, we again verify that the correlator is a superposition of on-shell scattering amplitudes.

Returning to  \Eq{eq:jO_simp}, we take the $\Delta\rightarrow 1$ limit that corresponds to the Milne soft limit $\omega\rightarrow 0$ that defines a massless vector in $\AdS_3$.  However, this limit requires care because the integral over $s$ is dominated near $s=0$ from infrared divergence in the amplitude.  In particular, the Weinberg soft theorem says that
\eq{
 \langle \textrm{out} |\epsilon_\mu \tilde J^\mu(sk) |\textrm{in} \rangle \overset{s\rightarrow 0}{=} \frac{1}{s} \sum_{i=1}^n  q_i \left(\frac{\epsilon k_i}{k k_i}\right)  \langle \textrm{out} |\textrm{in} \rangle + \textrm{regular in }s.
}
However, this $1/s$ singularity is regulated by oscillatory contributions coming from the $s^{\Delta-1}$ factor in the integrand, so
\eq{
\int_0^\infty \frac{ds}{s}\, s^{i\omega} (\cdot) = -\frac{i}{\omega} (\cdot) + \textrm{regular in }\omega .
}
The singularity in $\omega$ is cancelled by the prefactor in \Eq{eq:jO_simp}, which is proportional to $\omega$ in this limit.  Combining all terms, we then find that \Eq{eq:jO_simp} simplifies to the Weinberg soft factor in \Eq{eq:softgauge}, just as advertised.  Hence, we learn that the Milne soft limit $\omega\rightarrow 0$ and the Minkowski soft limit $E\rightarrow 0$ coincide, both generating the Weinberg soft theorem.

\subsection{Kac-Moody Algebra of $\CFT_2$} 
  
 \label{sec:KM} 
  
\begin{figure}[t]
\begin{center}
\includegraphics[width=.65\textwidth]{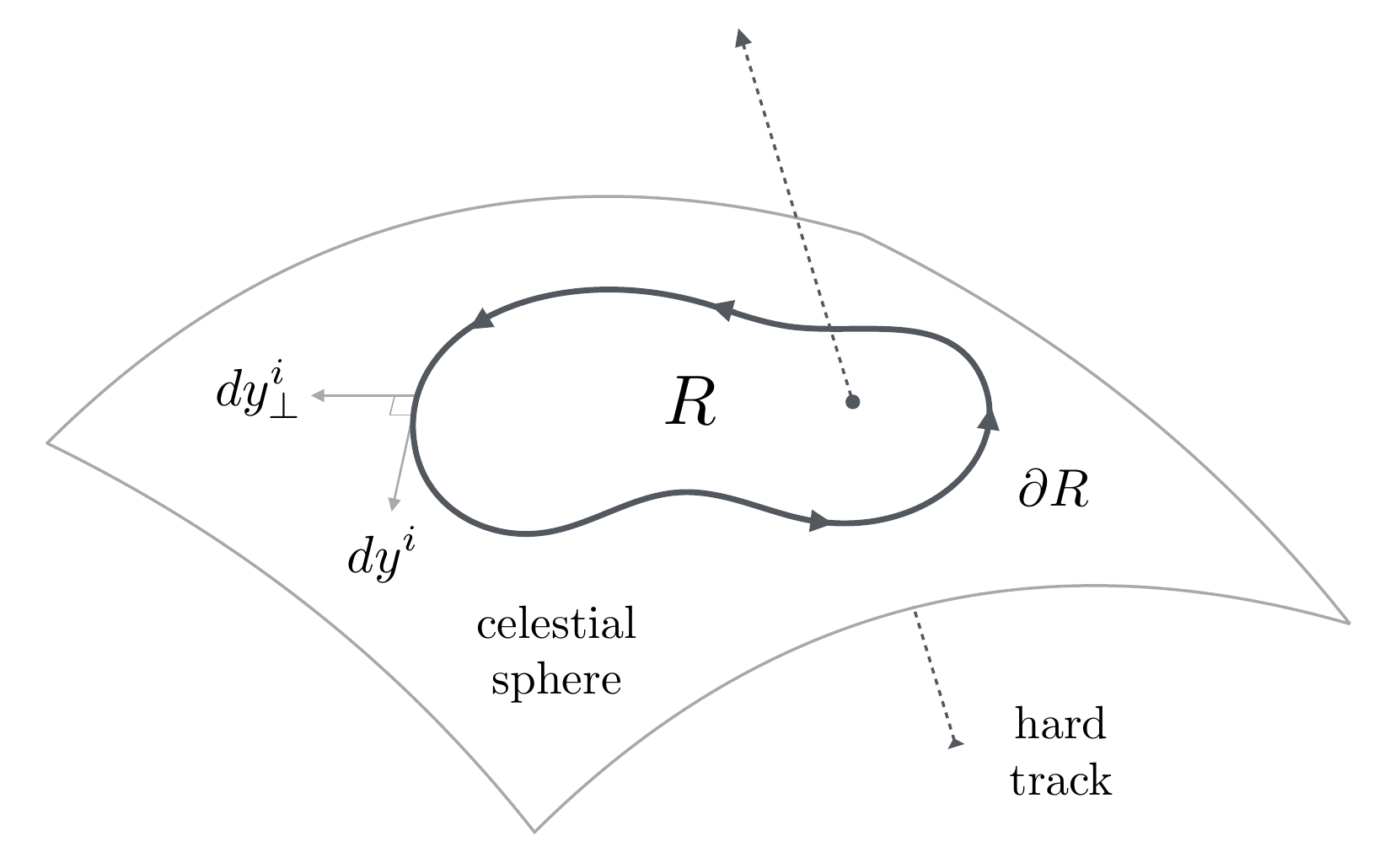}
\end{center}
\vspace*{-.5cm}
\caption{The celestial sphere houses a region $R$ whose boundary $\partial R$ encircles the trajectory of a hard particle.  The single helicity Aharonov-Bohm phase around $\partial R$ is simultaneously {\it i}) the cumulative charge of hard tracks threading $R$, {\it ii}) the integrated velocity kick experienced by test charges along $\partial R$, {\it i.e.}~the electromagnetic memory effect, and  {\it iii}) the Ward identity for the holomorphic conserved current of the 2D CFT.
Here $dy^i$ is the infinitesimal vector tangent to $\partial R$ while $dy_\perp^i$ is the infinitesimal vector orthogonal to $\partial R$ but still on the celestial sphere.  } 
 \label{fig:AB}
\end{figure}

The existence of a holomorphic conserved current $j(z)$ signals an infinite-dimensional symmetry algebra encoded in the $\CFT_2$ \cite{bigYellowBook}.  Since $\partial_{\bar z}j(z)=0$, we can Laurent expand the holomorphic current  in the usual fashion, 
\begin{equation}
j(z) = \sum_{m=-\infty}^\infty \frac{j_m}{z^{m+1}},
\end{equation}
yielding the infinitely many charges $j_m$ of an abelian Kac-Moody algebra.  Furthermore, a generalized ``soft charge'' can be defined with respect to a contour $\partial R$ in the $z$ coordinate bounding a 2D ``patch'' $R$ on the celestial sphere.  Such a patch is depicted in \Fig{fig:AB}. We can associate to this patch an arbitrary holomorphic function $\lambda(z)$ to define the soft charge,
\begin{equation}
j_{R,\lambda} = \oint_{\partial R} dz  \, \lambda(z) j(z).
\end{equation}
By the Ward identity for the 2D conserved current in \Eq{eq:softgauge} and Cauchy's theorem, this quantity counts number of charged particles in the scattering amplitude  threading the region $R$,
\begin{equation}
\langle  j_{R,\lambda} {\cal O}(z_1,\bar z_1)\cdots {\cal O}(z_n,\bar z_n) \rangle = \sum_{i \in R} q_i \lambda(z_i) \langle {\cal O}(z_1,\bar z_1)\cdots {\cal O}(z_n,\bar z_n) \rangle.
\end{equation}
This is an angle dependent charge conservation equation, where the left-hand side is the correlator of the ``soft charge'' and the right-hand side consists of the sum over hard particle charges within some angular acceptance.

Since $j(z)$ is a holomorphic current,  $\partial_{\bar z}$ acting on its correlators should vanish everywhere except at the insertion points of operators.  This is verified by applying $\partial_{\bar z}$ to \Eq{eq:softgauge}, yielding
\eq{
\partial_{\bar z} \langle j(z) {\cal O}(z_1,\bar z_1)\cdots {\cal O}(z_n,\bar z_n) \rangle = 2\pi \sum_{i=1}^n q_i\langle {\cal O}(z_1,\bar z_1)\cdots {\cal O}(z_n,\bar z_n) \rangle,
\label{eq:dj}
 }
where we have used the identity from complex analysis,
 \eq{
\partial_{\bar z}\left( \frac{1}{z} \right) = 2\pi \delta^2(z,\bar z).
\label{eq:d1z}
}
According to \Eq{eq:dj}, global charge conservation then requires that $\sum_{i=1}^n q_i =0 $, so the sum of all charges is zero.

\subsection{Chern-Simons Theory and Multiple Soft Emission}

\label{sec:gaugeCS}

In the previous sections we verified that soft gauge bosons in $\Mink_4$ correspond to massless vectors in $\AdS_3$  dual to conserved currents in a $\CFT_2$.  At the same time, we noted that the associated bulk-boundary propagators are pure gauge, suggesting an underlying $\AdS_3$ theory with no propagating degrees of freedom.  As this is the calling card of a topological gauge theory, CS theory is the natural candidate to describe the massless vectors of $\AdS_3$.  In this section we argue that this is precisely the case. We stress that the purely topological character is restricted to just the soft gauge sector of the 4D theory, dual to the 2D current algebra of the $\CFT_2$. More generally, the KK reduced $\AdS_3$ description is a CS gauge theory {\it coupled to non-topological matter}.  These degrees of freedom correspond to all 4D fields that carry finite Milne energy $\omega$.

\subsubsection{Abelian Chern-Simons Theory}

To begin, let us revisit the lifted bulk-boundary propagator $K_\mu$ as a solution to the classical field equations for a gauge field.  
Since the bulk-boundary propagator is pure gauge, its associated field strength vanishes everywhere, including on any $\AdS_3$ slice,
\eq{
\partial_i K_j-\partial_j K_i=0.
}
Rather trivially, this coincides with the equation of motion for an abelian CS gauge field $A_i$, whose field strength satisfies
\eq{
F_{ij}=0,
\label{eq:Fzero}
}
indicating the absence of propagating degrees of freedom expected in a topological theory.  Hence, far from sources, the bulk-boundary propagator $K_i$ is a solution to the equations of motion for a CS gauge field $A_i$, whose action is 
\eq{
S_{\rm CS} = \int_{\AdS_3} d^3 y\,   A_i F_{jk} \,  \varepsilon^{ijk}.
}
Since the CS theory is topological, the bulk spacetime, $\AdS_3$, is not so important, but the boundary, $\dAdS_3$, is crucial.  In fact, we must fix specific boundary conditions for the CS gauge theory.  Because the CS theory has a first order equation of motion, we can either specify $A_z$ on $\partial\AdS_3$ or $A_{\bar{z}}$ on $\partial\AdS_3$, but not both \cite{seibergMoore}. As we will soon see, these choices correspond to the soft $(+)$ or $(-$) helicity sectors of the 4D gauge theory, respectively.

It is instructive to see how this CS theory arises arises from the Milne soft limit, starting from the regime of finite Milne energy $\omega \neq 0$, where the scaling dimension is $\Delta =1 \pm i \omega$ according to \Eq{eq:Deltadef}.  As before, we interpret the lifted bulk-boundary propagator as a classical  gauge field solution,
$A_\mu = K^{\Delta}_\mu$.    It is easily checked that the associated field strength $F_{\mu\nu} \neq0$ for $\Delta\neq 1$ and therefore is 
not pure gauge.  However, the field strength satisfies the self-dual equation,
\eq{
F_{\mu\nu} = i \tilde F_{\mu\nu},
\label{eq:selfdual}
}
where the Hodge dual field is
\eq{
\tilde F_{\mu\nu} =\tfrac{1}{2}\varepsilon_{\mu\nu\rho\sigma} F^{\rho\sigma}.
} We now recall that the self-dual condition in \Eq{eq:selfdual} simply indicates that the electric and magnetic fields are phase shifted, consistent with a polarized electromagnetic wave.  Thus the self-dual condition restricts to the gauge field to the $(+)$ helicity sector. Had we began instead with with the complex conjugate bulk-boundary propagator, we would have obtained the anti-self-dual condition that defines the $(-)$ helicity sector. 

Note that for real gauge fields, self-duality is of course only possible in Euclidean signature.  However, we are in Lorentzian signature, so the self-dual condition implicitly entails a formal complexification of the gauge fields.   
  
In Milne coordinates, the  self-dual condition becomes
\eq{
F_{ij} = \tfrac{1}{2}i \epsilon_{ijk} \partial_\tau A_k,
\label{eq:Fijzero}
}
where we have dropped a term using the temporal Milne gauge condition $A_\tau=0$. Fourier transforming to Milne harmonics, we see that the right-hand side is proportional to the Milne energy, $i\partial_\tau =\omega$.  \Eq{eq:Fijzero} is then none other than the Proca-CS equation of motion for a gauge field of mass $i \omega$.
To revert to the case of a $\Delta=1$ conserved current, we take the corresponding limit of vanishing Milne energy $\omega=0$, in which case the right-hand side vanishes, reproducing our expression from \Eq{eq:Fzero}.  

 From the above analysis we conclude that the Witten diagrams corresponding to correlators of conserved $\CFT_2$ currents $j(z)$ and $\bar j(\bar z)$ are computed with $\AdS_3$ CS gauge fields describing soft gauge bosons of a single helicity in $\Mink_4$.   Importantly, our discussion thus far has centered on the abelian field equations, which automatically linearize so as to factorize the $(+)$ and $(-)$ helicity sectors.  In these theories the $(+)$ and $(-)$ helicity gauge bosons do not couple directly, so the corresponding $\CFT_2$ has both a conserved holomorphic current $j(z)$ and a conserved anti-holomorphic current $\bar j(\bar z)$.

\begin{figure}[t]
\begin{center}
\includegraphics[width=.9\textwidth]{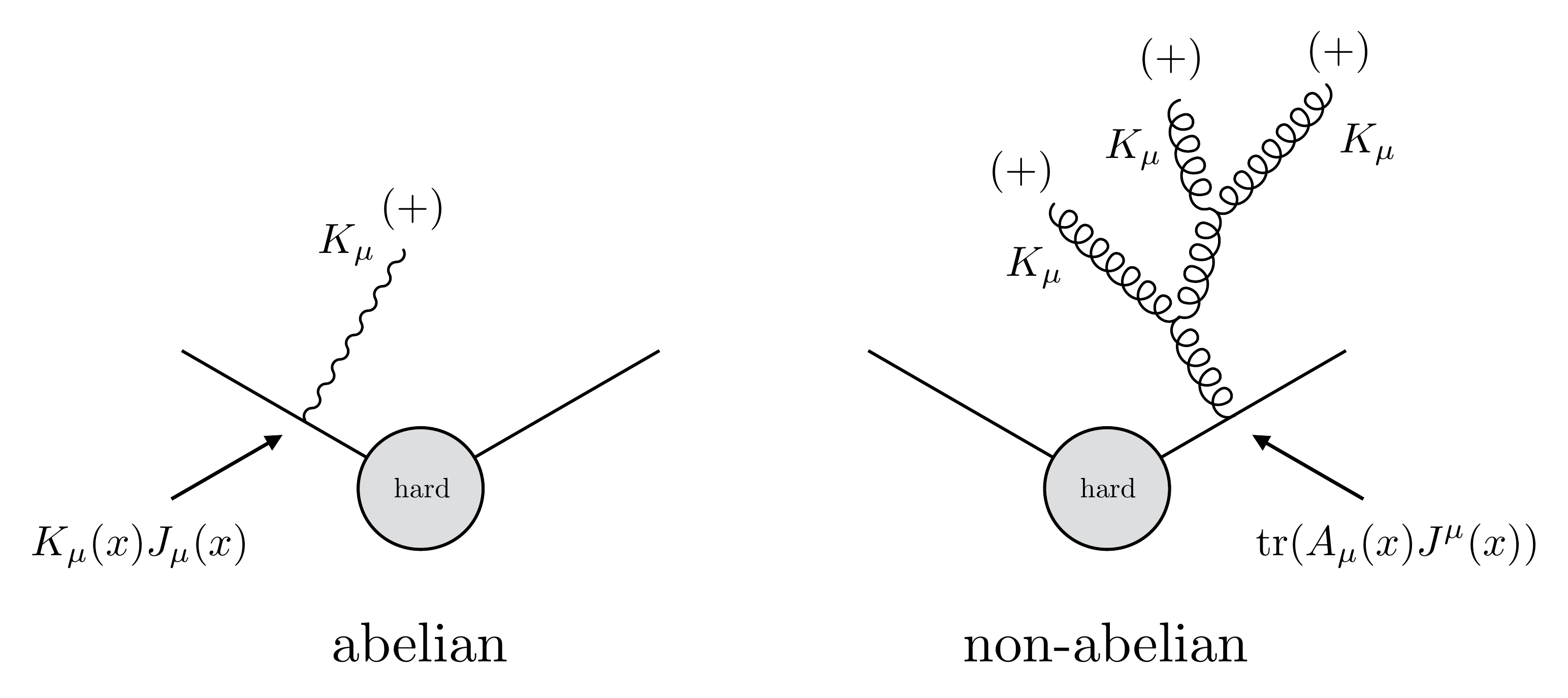}
\end{center}
\vspace*{-.5cm}
\caption{ Single emission of an abelian gauge boson and multiple emission of non-abelian gauge bosons.  In both cases, external legs connect to bulk-boundary propagators $K_{\mu}$.  In the non-abelian case, these soft emissions accumulate into a soft branch described by the field $A_\mu$. }
 \label{fig:soft_emissions}
\end{figure}

Up until now we have focused solely on soft sector of the gauge theory, neglecting  all hard quanta that appear in the form of hard charged matter or hard gauge bosons.  However, this relates to a possible point of confusion, which is that the self-dual solutions just described are only solutions of the {\it source free} equations of motion.  Naively, in the presence of sources, this self-duality will be spoiled.  This is, however, not actually a problem once we remember that the bulk-boundary propagators are by definition solutions to the source free, homogeneous equations of motion.  This is obvious because $K_{\mu\nu}$ is simply a function of its end points and not any particular property of a current.  We can see this diagrammatically in \Fig{fig:soft_emissions}, which shows how the bulk-boundary propagator undergoes self-dual, free propagation before making contact with a hard source.

Indeed, from this picture it is straightforward to see how the CS field interacts with hard sources.  Recall the Witten diagram corresponding to soft gauge boson emission,
\eq{
 \int d^4x \, K^{}_{\mu}(x) W^{\mu}(x) = \int d^3 y \, \sqrt{g} \, K_{i}(y) \int d \tau \, W^i(\tau, y),
 \label{eq:Wittentrack}
}
where in  \Eq{eq:Wdef}, $W^i$ denotes remainder of the Witten diagram, 
\eq{
W^i(\tau,y) = \langle \textrm{out} |J^i(\tau,y) |\textrm{in} \rangle,
\label{eq:Widef}
}
computed from the matrix elements of current $J^i$.   On the right-hand side we have used the fact that in Milne coordinates, the bulk-boundary propagator is $\tau$ independent since it corresponds to a Milne zero mode.   Hence, we see that if $K_i$ is to be interpreted as a classical configuration of a CS gauge field $A_i$ in $\AdS_3$, then it couples to a Milne time-integrated version of $W^\mu$, given by
\eq{
W^i_{\rm eff}(y)  = \int d \tau \, W^i(\tau,y)  =  \langle \textrm{out} |J_{\rm eff}^i(y) |\textrm{in}\rangle ,
\label{eq:Wi_tracks}
}
where we have defined a Milne time-integrated current,
\eq{
J^i_{\rm eff}(y)  =  \int d \tau \, J^i(\tau,y).
}
If we think of the full current $J^\mu$ as physically representing an array of hard particle world lines and interactions in $\Mink_4$, then the Milne time-integrated current $J^i_{\rm eff}$ is a static record of the hard ``tracks'' defined by these trajectories throughout all of time.  It is to these hard tracks in $\AdS_3$ to which the CS gauge field $A_i$ couples.  See \Fig{fig:soft_emissions} for a schematic depicting the absorption of an abelian soft gauge boson by a hard track.

Altogether, we see that this describes a purely AdS$_3$ description derived from a ``KK reduction'' of hard particles in $\Mink_4$ into massive $\AdS_3$ fields coupled covariantly to the CS gauge field, $A_i$.  The corresponding action is then
 \eq{
 S_{\rm CS} &= \int_{\rm \AdS_3}  d^3 y \,  A_i F_{jk}  \varepsilon^{ijk} + 
 \int_{\rm \AdS_3}  d^3 y \, \sqrt{g} \, A_{i} {J}^i_{\rm eff}.
 \label{eq:CSaction}
 }
Massive $\AdS_3$ fields contribute to 3D Witten diagrams which are equivalent to the matrix elements of the 4D current $J^\mu$ projected down to the zero mode $J^i_{\rm eff}$ in order to couple to the CS gauge field describing the Milne soft mode.
So the couplings of the bulk-boundary propagator match to a CS action given by \Eq{eq:CSaction}.

\subsubsection{Non-Abelian Chern-Simons Theory}

For non-abelian gauge theories the story is more complicated because $(+)$ and $(-)$ helicity gauge bosons interact directly and non-linearly.  Nevertheless, a similar story applies.   To understand why, consider tree-level non-abelian gauge theory, subject to a restriction of the external states to be a single helicity, say $(+)$. Let us not even take the soft limit---instead, consider both soft and hard $(+)$ particles for the purpose of this discussion.

 By definition, the bulk-boundary propagators $K_\mu^{\Delta}$ satisfy the self-dual condition in \Eq{eq:selfdual}.  The non-abelian subtlety arises because multiple external soft gauge bosons will in general interact and merge into soft ``branches'' which then attach to the hard bulk current, as depicted in \Fig{fig:soft_emissions}.  Mathematically, each soft branch can be described by a non-abelian gauge field $A_\mu(x)$, defined from the corresponding Feyman diagram for that particular tree of soft gauge bosons.  So explicitly, $A_\mu(x)$ is some integral over products of soft interaction vertices and bulk-boundary and bulk-bulk propagators.  Here $x$ is the bulk point at which the soft branch connects to the hard diagram, again as indicated in \Fig{fig:soft_emissions}.  So by definition, $A_\mu(x)$ is comprised solely of soft elements.  
 
\begin{figure}[t]
\begin{center}
\includegraphics[width=.9\textwidth]{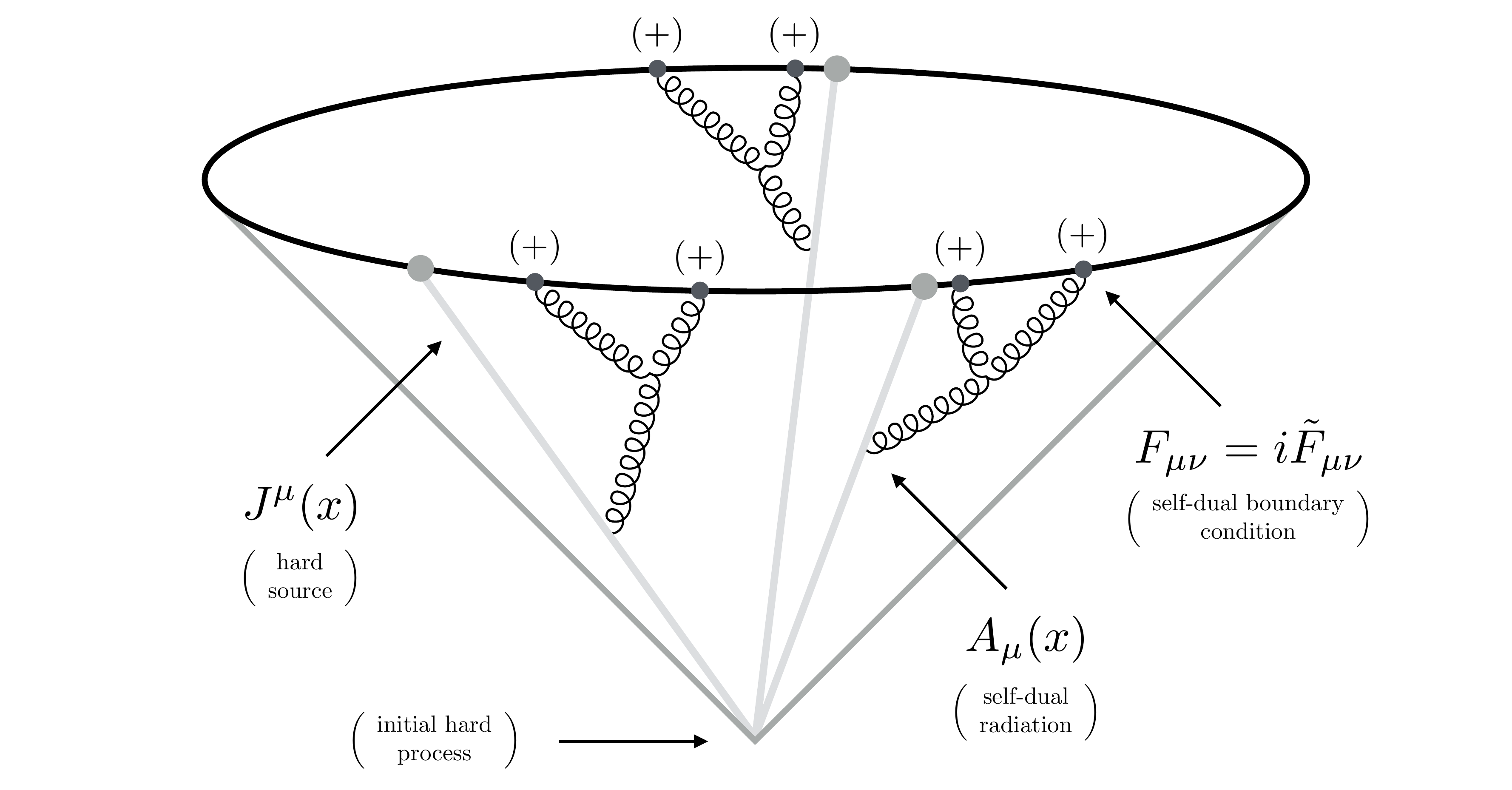}
\end{center}
\vspace*{-0.5cm}
\caption{Schematic depicting soft, single helicity non-abelian gauge bosons coupling to hard sources.    Each soft branch is initiated by a set of $(+)$ helicity soft gauge bosons, so the corresponding field configuration is self-dual.   }
 \label{fig:soft_radiation}
\end{figure}

 In this way we see that the soft branch field $A_\mu$ is just the perturbative expansion of a classical solution to the non-abelian YM equations of motion, where the free limit reverts to a superposition of $K_\mu^{\Delta}$ bulk-boundary propagators for the external lines. Since all external lines are taken to be $(+)$ helicity, this linear superposition is self-dual. In turn, this implies that the full non-linear soft branch $A_\mu$ is also non-linearly self-dual, since 
self-dual configurations continue to be self-dual upon non-linear classical evolution. 
 This follows since the self-dual equations are first order and thus guarantee satisfaction of the second order YM field equations.   
 As before, one can naively worry about violation of the self-dual condition by sources.  However, there is again no obstruction because the soft branch field is a solution to the source free non-linear equations of motion, independent of the hard source.  
A schematic of our physical picture is shown in \Fig{fig:soft_radiation}. 
 
We thereby conclude that the soft branch field $A_\mu$ satisfies the 4D non-abelian self-dual equations, given by \Eq{eq:selfdual} where $A_\mu$ is the matrix valued gauge field and and $F_{\mu\nu}= \partial_\mu A_\nu - \partial_\nu A_\mu -[A_\mu,A_\nu]$ is the full non-linear field strength.  Moreover, in Milne temporal gauge, the non-abelian CS equations of motion are given by \Eq{eq:Fijzero} where the left-hand side contains the full non-linear field strengths. By momentum conservation, if the external legs of the soft branch are Milne soft, then so too is $A_\mu$, so the right-hand side of \Eq{eq:Fijzero} is zero.  Thus, we verify that the soft branch field satisfies the non-abelian CS equation of motion.

Finally, let us discuss the interactions of the non-abelian CS fields with the hard process.  The analysis is same as for the abelian case, except the hard process couples to non-abelian soft branches rooted in a multiplicity of soft external gauge bosons, rather than a single abelian bulk-boundary propagator.  In particular, the Witten diagram for multiple soft emission is
\begin{equation}
\int d^4x \, {\rm tr} (A_{\mu}(x) W^{\mu}(x)) = \int d^3 y \, \sqrt{g} \, {\rm tr} ( A_{i}(y)  W^i_{\rm eff}(y)), 
\end{equation}
where $A_\mu$ is the soft branch field and $W_\mu$ again characterizes the hard current.  Here we have defined a Milne time-integrated current $W^i_{\rm eff}$ as in \Eq{eq:Wi_tracks}, only for a matrix valued current.  Reminiscent of KK reduction, we see that the hard particles in Minkowski space couple to the soft branch field only through a zero mode projection of the hard current.  

Finally, the Witten diagram associated with a non-abelian gauge field is also \Eq{eq:Wittentrack}, only with matrix valued gauge fields and a color trace.  In turn, this implies that multiple soft gauge boson emissions are dictated by a non-abelian CS action,
 \eq{
 S_{\rm CS} &= \int_{\rm \AdS_3}  d^3 y\,   {\rm tr}\left(A_i F_{jk}  + \frac{2}{3}A_i A_j  A_k \right)  \varepsilon^{ijk} + 
 \int_{\rm \AdS_3} d^3 y \, \sqrt{g} \,  {\rm tr} \left( A_{i} {J}_{\rm eff}^i \right),
 }
where as before, we have defined ${J}_{\rm eff}^i$ to be the Milne time-integrated ``tracks'' of the hard particles in the scattering process. 

As in the abelian case, the first order nature of the CS theory requires that we specify a boundary condition for $A_z$ or $A_{\bar z}$ but not both.  We see that this corresponds to keeping a single helicity in the soft limit.  
This explains the proposal of \cite{stromingerKacMoody} to restrict to single helicity soft limits because of the non-commutation of opposite helicity soft limits in non-abelian gauge theory.  This contrasts with abelian gauge theory, where both helicities can be described simultaneously because they do not interact which each other directly.   

\subsubsection{Locating Chern-Simons Theory in $\Mink_4$}

In the previous sections we constructed abelian and non-abelian CS theories characterizing multiple emissions of soft, single helicity gauge bosons.  The CS gauge fields interact with a Milne time-integrated current describing the tracks of hard particles.  While the underlying 3D spacetime is the $\AdS_3$ obtained by dimensional reduction, it will be illuminating to understand where the CS gauge field is precisely ``located'' in 4D spacetime.  To see this we now consider a slightly different but more intuitive derivation.

For simplicity, consider the case of abelian gauge theory, where we solve the field equations in the presence of a current.  This differs from our earlier approach, where soft branches were described by a gauge field satisfying the source free equations of motion.  Here we instead {\rm start} with the current source and then compute the resulting gauge field configuration, in line with the usual approach taken in classical electrodynamics.

In Milne temporal gauge and working in $\omega$ frequency space, the gauge field generated by a particular current is
\begin{equation} 
A_i (\omega, y) = \int d^3y' \, \sqrt{g}  \, G_{ij}(\omega, y,y')J^j(\omega, y'), 
\label{eq:maxwellSolution}
\end{equation}
where the right-hand side is the current convolved with a Proca propagator satisfying \Eq{eq:procaEq} for a vector in $\AdS_3$ of  ``mass squared'' equal to $- \omega^2$. 

The key observation is that the Proca wave equation for a vector in $\AdS_3$ factorizes \cite{townsend} into 
\eq{
\nabla_k \nabla^k\delta_i^{\;j} - \nabla_i \nabla^j +  \omega^2 \delta_i^{\; j}=\Pi^{+k}_i \Pi^{-j}_k=\Pi^{-k}_i \Pi^{+j}_k,
\label{eq:factorize}
}
where each projection operator is 
\eq{
\Pi^{\pm j}_i &= \omega \delta_i^{\;j} \pm \epsilon_i^{\; jk} \nabla_k.
}
From the form of \Eq{eq:factorize}, it is clear that any functions that are annihilated by $\Pi^{\pm j}_i$ will also be annihilated by the wave equation.  \Eq{eq:factorize} then implies that the Proca propagator is
\eq{
\label{eq:disaggregate}
G_{ij} = \frac{G^+_{ij} + G^-_{ij}}{2 \omega},
}
where $G^{\pm}_{ij}$ separately satisfy the first order wave equations,
\eq{
\Pi^{\pm j}_i G^\pm_{jk} = i\delta_{ik}\delta^3(y, y').
}
These are nothing more than the equations of motion for a pair of Proca-CS fields of ``mass'' $\pm i \omega$, so $G^{\pm}_{ij}$ denote the corresponding Proca-CS propagators.  We thus find that an off-shell Proca vector in $\AdS_3$ can be described by a pair of off-shell Proca-CS fields. 

Now let us return to \Eq{eq:maxwellSolution}, taking the matrix element of this equation between 4D in and out states.  Using \Eq{eq:Widef}, this sends the current $J^i$ to the quantity $W^i$, which characterizes hard particles sources.  \Eq{eq:maxwellSolution} then becomes
\begin{equation}
\langle  {\rm out} | A_i (\omega,y)| {\rm in} \rangle = \int d^3y' \, \sqrt{g}  \, G_{ij}(\omega, y,y') W^j(\omega, y').
\end{equation}
Because the bulk-bulk propagator splits into halves as discussed before, so too does the resulting gauge field $A_i$,
\begin{equation} \label{eq:CSprocaField}
\langle  {\rm out} | A^\pm_i (\omega,y)| {\rm in} \rangle = \frac{1}{2 \omega} \int d^3y' \, \sqrt{g}  \, G^{\pm}_{ij}(\omega, y,y') W^j(\omega, y').
\end{equation}
For values of $y$ in the bulk of $\AdS_3$, the physical significance of these off-shell Proca-CS fields is not completely transparent.  However, as $y$ approaches the boundary of $\AdS_3$, the gauge field $A_i$ becomes radiation-dominated and $A_i^\pm$ should be interpreted as the two helicities of on-shell electromagnetic radiation.  Since the bulk-boundary propagator is just the boundary limit of the bulk-bulk propagator, we see that we have just been computing the same Witten diagram for gauge boson emission discussed in our earlier derivation of the CS structure.  In any case, away from the boundary,
the Proca-CS field of \Eq{eq:CSprocaField} can be understood as an off-shell extension of the helicity decomposition for general $y$. 

\begin{figure}[t]
\begin{center}
\includegraphics[width=.8\textwidth]{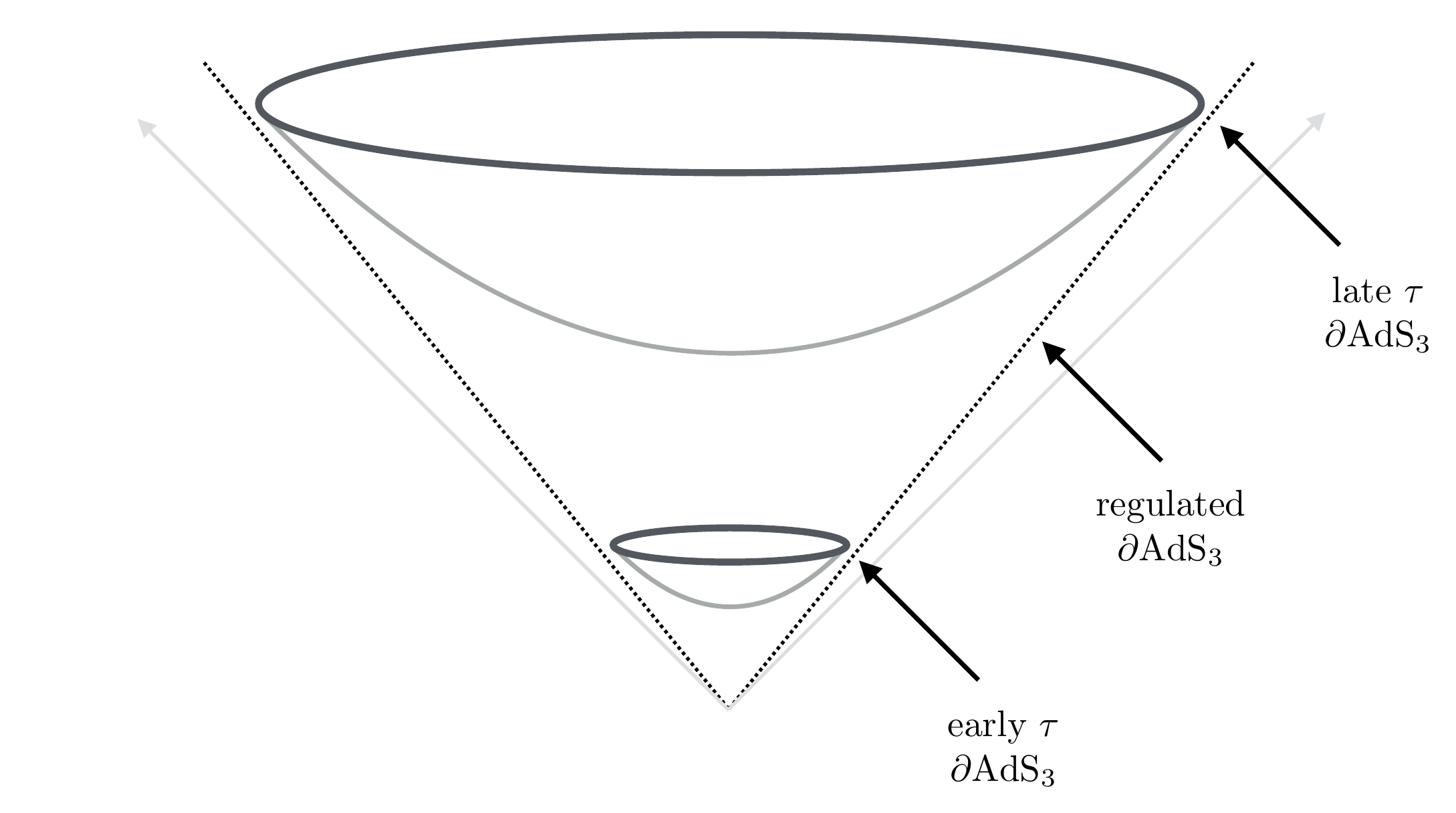}
\end{center}
\vspace*{-.75cm}
\caption{Depiction of the regulated boundary $\dAdS_3$.   At late Milne times $\tau\rightarrow +\infty$, the boundary of the correspoding $\AdS_3$ slice   approaches null infinity.  Meanwhile, at early Milne times $\tau \rightarrow -\infty$, this boundary approaches the origin.  }
 \label{fig:regulated_boundary}
\end{figure}

To explicitly construct the CS gauge field we simply take the Milne soft limit of the Proca-CS field.  However, from \Eq{eq:CSprocaField} it is clear that this limit only exists if we first multiply by $\omega$.  Thus, the CS gauge fields must correspond to the modified limit,
\begin{equation} \label{eq:CSfield}
 -i \lim_{\omega \rightarrow 0} \omega A_i^{\pm}(\omega, y) =  \int d \tau \, \partial_{\tau} A_i^{\pm}(\tau, y) = A_i^{\pm}(\tau \rightarrow + \infty, y) 
- A_i^{\pm}(\tau \rightarrow - \infty, y). 
\end{equation}  
We can better understand this result by ``regulating'' the boundary of $\AdS_3$ in the standard way used in the context of AdS/CFT in global coordinates.  In terms of the 4D embedding, this prescription corresponds to an infinitesimal ``narrowing'' of the lightcone bounding the Milne wedge, as depicted in \Fig{fig:regulated_boundary}.  
We then see that within the regulated Milne region, $\tau \rightarrow - \infty$ is the Minkowski origin, so the soft field is trivial there.

We thereby conclude that from the 4D perspective, the CS gauge field is the {\it single helicity and Milne late} soft field at null infinity.  At the same time, the charged matter of the CS theory is captured by $J^i_{\rm eff}$, which characterizes the hard-particle tracks in the bulk of Minkowski, cumulative over all time. We will see later how this relates to the phenomenon of electromagnetic memories.

\subsection{Wess-Zumino-Witten Model and Multiple Soft Emission}

\label{sec:WZW}

We can now invoke the established $\AdS_3/\CFT_2$ dictionary to relate our CS theory in $\AdS_3$ to the chiral half of a rational $\CFT_2$ known as the WZW model \cite{wittenJones,seibergMoore,wittenHolomorphic,stromingerCSAdSCFT}.  The WZW model is simply a 2D non-linear sigma model defined by the action, 
\eq{
S_{\rm WZW} = \int_{\partial \AdS_3} d^2 z\, {\rm tr} (\partial_{\bar z} U^{-1} \partial_z U)  + \frac{1}{3} \int_{\AdS_3} d^3y\,    {\rm tr} (\partial_i U U^{-1} \partial_j U U^{-1} \partial_k U U^{-1})\,\varepsilon^{ijk},
}
where the second term is topological.  As is well-known \cite{bigYellowBook}, the WZW model enjoys both a holomorphic current $j^a(z)$ and an anti-holomorphic current $\bar j^a(\bar z)$. Laurent expanding $j^a(z)$ gives
\eq{
j^a(z) = \sum_{m=\infty}^\infty  \frac{j_m^a}{z^{m+1}},
\label{eq:KMgenerator}
}
where $j_m^a$ are the generators of an infinite-dimensional non-abelian Kac-Moody algebra. Notably, CS theory is equivalent to just the holomorphic sector of the WZW model \cite{wittenJones,seibergMoore,wittenHolomorphic,stromingerCSAdSCFT}, matching with the fact that it only describes a single helicity of the 4D gauge theory.

The operator product expansion for a conserved current in the WZW model is
\eq{
j^a (z) {\cal O}^b(z')\sim \frac{f^{abc} {\cal O}^c(z')}{z-z'},
}
which from \Eq{eq:softgaugeNA} is plainly equivalent to the Weinberg soft factor for a non-abelian gauge theory.  As $z$ denotes stereographic coordinates on the celestial sphere, we see that the operator product expansion is dual to an expansion in the collinear singularities of scattering amplitudes.

Let us comment on an innocuous but perhaps important fact about the WZW model, which is that the stress tensor is directly related to the holomorphic current via the Sugawara construction \cite{bigYellowBook},
\eq{
t(z) \sim \sum_a j^a(z) j^a(z).
}
The existence of a stress tensor for the current algebra alone reinforces the fact that the soft sector described by the WZW model is itself a consistent  sub-$\CFT_2$ within the full $\CFT_2$ describing both soft and hard particles. In the sub-$\CFT_2$, the hard particles are only visible as soft color sources represented by Wilson lines along the hard tracks.  However, the dynamics of the hard particles themselves require additional structure which will add additional contributions to the stress tensor beyond the Sugawara construction.

At the same time,  the Sugawara construction implies a connection between the 2D stress tensor and double collinear gauge boson emission.  
This structure is highly suggestive given the known link between the self-dual sectors of gauge theory and gravity \cite{donalOConnell} which manifests the so-called BCJ double copy \cite{BCJ}.

\subsection{Relation to Memory Effects}

\label{sec:gaugememory}

We have argued that 4D scattering amplitudes for soft gauge boson emission are described by 3D CS gauge theory with matter.
To the soft sector, the hard particles appear as Wilson line color sources along their tracks. These Wilson lines, together with insertions of the 2D conserved current operators formulate a classic CS calculation that reproduces the known 4D soft factors.

A corollary of this CS structure is an intrinsic topological character to 4D soft emission.  Consistent with this picture, we will see how the soft sector elegantly exhibits the physics of the abelian and non-abelian AB effects---the hallmark of CS physics \cite{wittenJones,nonAbelianAB1,nonAbelianAB2}. For simplicity, we focus for now on the abelian case, identifying these AB effects and relating them to the previously identified phenomenon of electromagnetic memories \cite{qedMemory,pasterskiQEDmemory,susskindQEDmemory}.

\subsubsection{Chern-Simons Memory and the Aharonov-Bohm Effect}

To begin, let us consider the contour-integrated form of the $\CFT_2$ Ward identity for the holomorphic current derived in \Eq{eq:softgauge}, 
\eq{
\oint_{\partial R} dz\,  \langle j(z) {\cal O}(z_1,\bar z_1)\cdots {\cal O}(z_n,\bar z_n)  \rangle =\sum_{i \in R} q_i  \langle  {\cal O}(z_1,\bar z_1)\cdots {\cal O}(z_n,\bar z_n)  \rangle,
\label{eq:intWard}
 }
where again $R$ is a 2D patch on the celestial sphere near the boundary of $\AdS_3$.  This region is depicted in \Fig{fig:AB}.
From our earlier discussion, we saw that the current algebra of the $\CFT_2$ is dual to a CS theory describing soft, single helicity gauge bosons in $\Mink_4$ at late Milne time $\tau \rightarrow + \infty$.  By the standard $\AdS_3/\CFT_2$ grammar applied to CS,
\eq{
j(z) \sim A_z(\tau\rightarrow +\infty,\rho \rightarrow 0,z),
}
where $\sim$ denotes the holographic duality and we implicitly take the limit $\tau\rightarrow +\infty$ before $\rho\rightarrow 0$.
Here $A_z$ is simply $z$ component of the soft gauge field on $\dAdS_3$ at late Milne times.   Plugging back into \Eq{eq:intWard} we obtain
\eq{
\oint_{\partial R} dz  \, A_z(\tau\rightarrow +\infty, \rho\rightarrow 0, z)  =\sum_{i \in R} q_i,
\label{eq:wardAdS}
 }
which is implicitly evaluated inside a correlator with additional hard operators, as in \Eq{eq:intWard}.  To avoid unnecessary notational clutter, this will also be true of the rest of the expressions in this section. 

The above result has the form of a 3D AB phase for the CS gauge field at $\tau \rightarrow +\infty$.  From our earlier analysis, we saw that the equation of motion for the CS gauge field in the presence of hard sources is
\begin{equation} 
F_{ij} = \epsilon_{ijk} {J}^k_{\rm eff},
\end{equation}
where ${J}^i_{\rm eff} = \int d\tau J^i$ is the Milne time-integrated current.    Integrating this equation over the region $R$, we find that
\begin{equation}
\int_R F = \int_R *{J}_{\rm eff} = \int d \tau \int_R *J = \sum_{i \in R} q_i.
\end{equation}
Therefore, the AB phase around the loop $\partial R$ is equal to the 
the field strength flux through $R$, which is in turn equal to the Milne time-integrated charge flux through $R$.
In this way the hard particles in the scattering process will induce AB phases in the CS gauge field characterizing soft, single helicity emissions.

It is important to realize that the AB phase under discussion is {\it not} literally the standard 4D AB effect, but rather a 3D ``chiral'' version restricted to single helicity radiation.  In particular, the complex contour integral performed in \Eq{eq:wardAdS} only picks out the $(+)$ helicity component,
\eq{
\oint_{\partial R} dz \,  A_z(\tau\rightarrow +\infty,y)  = \oint_{\partial R} d y^i  \, A^{+}_i(\tau \rightarrow + \infty, y) .
 }
The restriction to a single helicity is crucial---without it we would have
\eq{
\oint_{\partial R} d y^i \, A_i(\tau \rightarrow +\infty, y)= \oint_{\partial R} d y^i \, A^{+}_i(\tau \rightarrow + \infty, y) + 
A^{-}_i(\tau \rightarrow + \infty, y) = \int_R  F^+ + F^- = 0.
\label{eq:cancel}
 }
The last expression is the integral over $R$ of the total field strength, including both $(+)$ and $(-)$ contributions.  Since $R$ lies on the celestial sphere, the integral runs over the {\it radial} magnetic flux at null infinity which vanishes due to the transversality of asymptotic electromagnetic radiation. 

The physical interpretation of \Eq{eq:wardAdS} becomes more transparent if we realize that the restriction to $(+)$ helicity modes indirectly relates the components of the gauge field tangent to the contour to the components {\it normal} to the contour but still tangent to the celestial sphere, as shown in \Fig{fig:AB}.   Applying this also to the $(-)$ helicity components together with the cancellation in \Eq{eq:cancel}, we find 
\begin{equation}
\oint_{\partial R} d y^i \, A^{\pm}_i(\tau \rightarrow +\infty, y)  = \frac{1}{2} \oint_{\partial R} d y^i_{\perp}  A_i(\tau \rightarrow + \infty, y),   
\end{equation}
where $dy^i_\perp$  is the vector perpendicular to $dy^i$ but tangent to the celestial sphere.  Crucially, the $A_i$ on the right-hand side of the above equation is not restricted by helicity, so $A_i = A^+_i + A^-_i$.  

A simple physical interpretation of the above result follows if we consider a scattering process for a set of electrically neutral in states scattering into charged out states.  In this case we are permitted to restrict to the regulated Milne region of \Fig{fig:regulated_boundary}, in which case $A_i(\tau\rightarrow -\infty,y)=0$ as an initial condition.  The above equation then becomes
 \begin{equation}
\oint_{\partial R} d y^i \,  A^{\pm}_i(\tau \rightarrow +\infty, y) = \frac{1}{2} \oint_{\partial R} d y^i_\perp \int d \tau \,  E_i(\tau, y) ,   
\end{equation}
where $E_i = \partial_{\tau} A_i$ is the electric field. A massive probe charge undergoes acceleration proportional to the local electric field, so the ``memory field''
$\int d \tau  E_i$ is literally equal to the cumulative ``velocity kick'' received by the probe from the soft radiation in the scattering process. 
This is nothing but the electromagnetic memory effect \cite{qedMemory}.  

We have shown that the AB phase is simply a contour of the memory field $\int d \tau  E_i$.  Consequently, one might then wonder if the AB phase somehow contains less information than the original memory field.  In fact, it is actually possible to construct the memory field from the AB phase since the naively missing information is given by the curl of the electric field.  However, by the classical field equations, this is proportional to $\partial_{\tau} B_i$, which vanishes in the Milne soft limit.  We thereby conclude that the AB phase and by extension the CS gauge field is equivalent to the memory field.  

\subsubsection{Chern-Simons Level from Internal Soft Exchange}

\begin{figure}[t]
\begin{center}
\includegraphics[width=.45\textwidth]{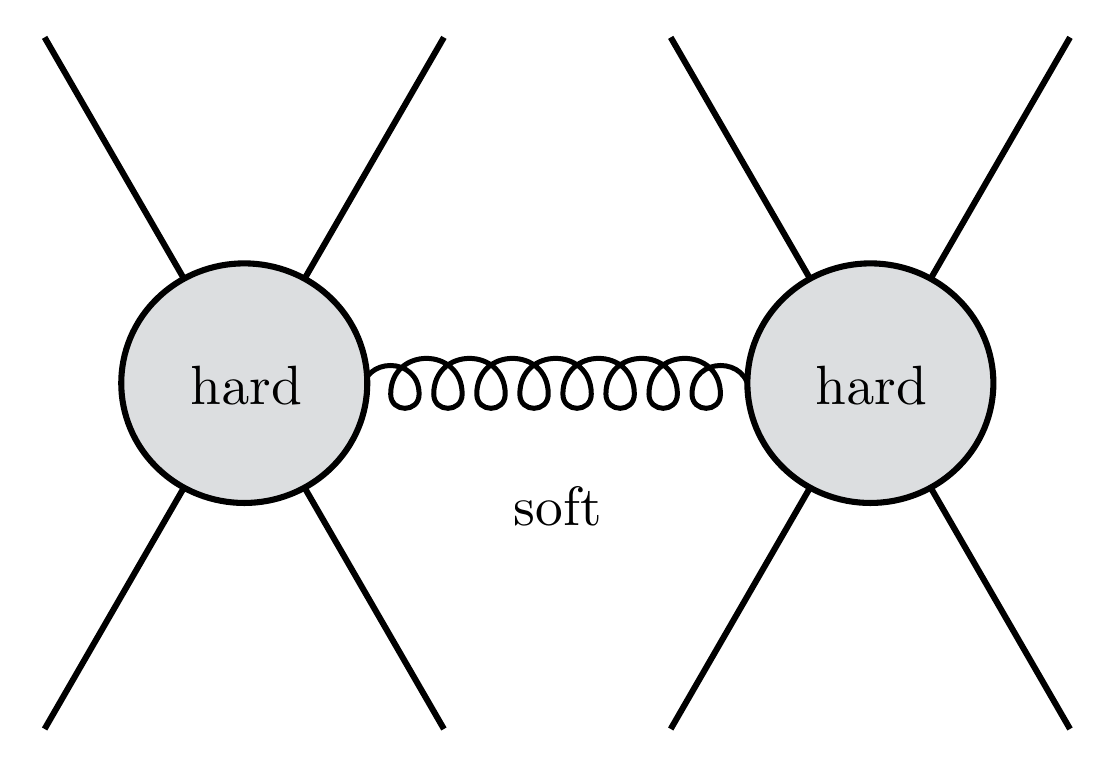}
\end{center}
\vspace*{-0.5cm}
\caption{Internal exchange of a soft gauge boson between two hard processes.  The associated Feynman diagram comes with a factor of $1/g_{\rm YM}^2$ while the associated Witten diagram comes with $k_{\rm CS}$, suggesting that $k_{\rm CS} \sim 1/g_{\rm YM}^2$.}
 \label{fig:soft_exchange}
\end{figure}

We have just seen how  the AB effect in the 3D CS description for 4D soft emission encodes a velocity kick for charged particles that embodies the electromagnetic memory effect.  While electromagnetic memory is most simply measured with massive charged probes, an alternative approach would be to configure a secondary hard process comprised massless charged particles that measure the soft emission from an initial scattering.    In the CS theory, this corresponds to diagrams composed of disjoint charged currents connected only by the exchange of an {\it internal} CS gauge line, as depicted
in \Fig{fig:soft_exchange}. This requires a new element, as thus far we have only matched the external CS lines to external soft emission lines in 4D.

Obviously, the exchange of an internal CS gauge field in $\AdS_3$ is dual to a $\Mink_4$ scattering amplitude with an internal soft gauge boson exchange.  Such an amplitude describes two hard processes connected by a soft internal gauge boson, so it only occurs at very special kinematics.  Since this particle travels a great distance before it is reabsorbed, it can be assigned a helicity.  The external soft emission and absorption processes studied earlier are then just sewn together as factorization channels of this composite process. 

Internal gauge exchange in CS is also important in because it encodes the CS level, $k_{\rm CS}$,  reflecting quantum fluctuations of the gauge field. When the Lie algebra is normalized independently of the couplings of the gauge theory, the CS action reads
 \eq{
 S_{\rm CS} &= \frac{k_{\rm CS}}{4\pi} \int_{\rm \AdS_3}  d^3 y\,   {\rm tr}\left(A_i F_{jk}  + \frac{2}{3}A_i A_j  A_k \right)  \varepsilon^{ijk},
 \label{eq:action_level}
 }
while the action for YM theory in 4D is
\begin{equation}
S_{\rm YM} =-\frac{1}{2g_{\rm YM}^2} \int_{\Mink_4} d^4 x \, {\rm tr} \left( F_{\mu \nu} F^{\mu \nu}\right).
\end{equation}
Notably, the solutions to the classical CS and YM equations of motion {\it do not depend} on $k_{\rm CS}$ nor $g_{\rm YM}$ since these are prefactors of the action, and thus drop out of the homogenous field equations.  Said another way, at tree level these couplings can be reabsorbed into the definition of $\hbar$.  Hence, the gauge field describing the soft external branches depicted in \Fig{fig:soft_radiation} are actually independent of these parameters.  On the other hand, these variables do enter into diagrams with internal CS gauge lines, or equivalently $\Mink_4$ processes with intermediate soft gauge boson exchange.  In CS perturbation theory \cite{CSpert1,CSpert2}, the former comes with a factor of $1/k_{\rm CS}$ and the latter, with a factor of $ g_{\rm YM}^2$. Therefore, we conclude that
\begin{equation} \label{eq:level}
k_{\rm CS} \sim \frac{1}{g_{\rm YM}^2},
\end{equation} 
in agreement with \cite{stromingerFirstPaper} but not \cite{stromingerKacMoody}, which argued for a vanishing Kac-Moody level. 

This result can also be obtained from the following heuristic derivation.  Substituting the self-dual constraint, $F_{\mu\nu} = i \tilde{F}_{\mu\nu}$, into the YM action in the regulated Milne region of \Fig{fig:regulated_boundary}, we find that
\eq{
S_{\rm YM} = -\frac{i}{2g_{\rm YM}^2} \int_{\rm \Mink_4} d^4x \, {\rm tr}(F_{\mu\nu} \tilde F^{\mu\nu}) &= \frac{i}{2g_{\rm YM}^2} \int_{\rm \Mink_4} d^4 x\, \partial_\sigma \, {\rm tr}\left(A_\mu F_{\nu\rho}  + \frac{2}{3}A_\mu A_\nu  A_\rho \right)  \varepsilon^{\mu\nu\rho\sigma},
}
which is a total derivative.
In principle, this total derivative will integrate to all the boundaries of the regulated $\Milne_4$.  However, due to our choice of Milne temporal gauge $A_\tau=0$ and the Milne soft limit $F_{\tau i} = 0$, the only boundary that contributes is at late $\tau$.  Thus, we again obtain the non-abelian CS action in \Eq{eq:action_level} where $A_i$ is the gauge field at $\tau\rightarrow +\infty$.  Matching this to the CS action, we verify \Eq{eq:level}.

\subsection{Toy Model for a Black Hole Horizon}

\label{sec:toymodel}

\begin{figure}[t]
\begin{center}
\includegraphics[width=.5\textwidth]{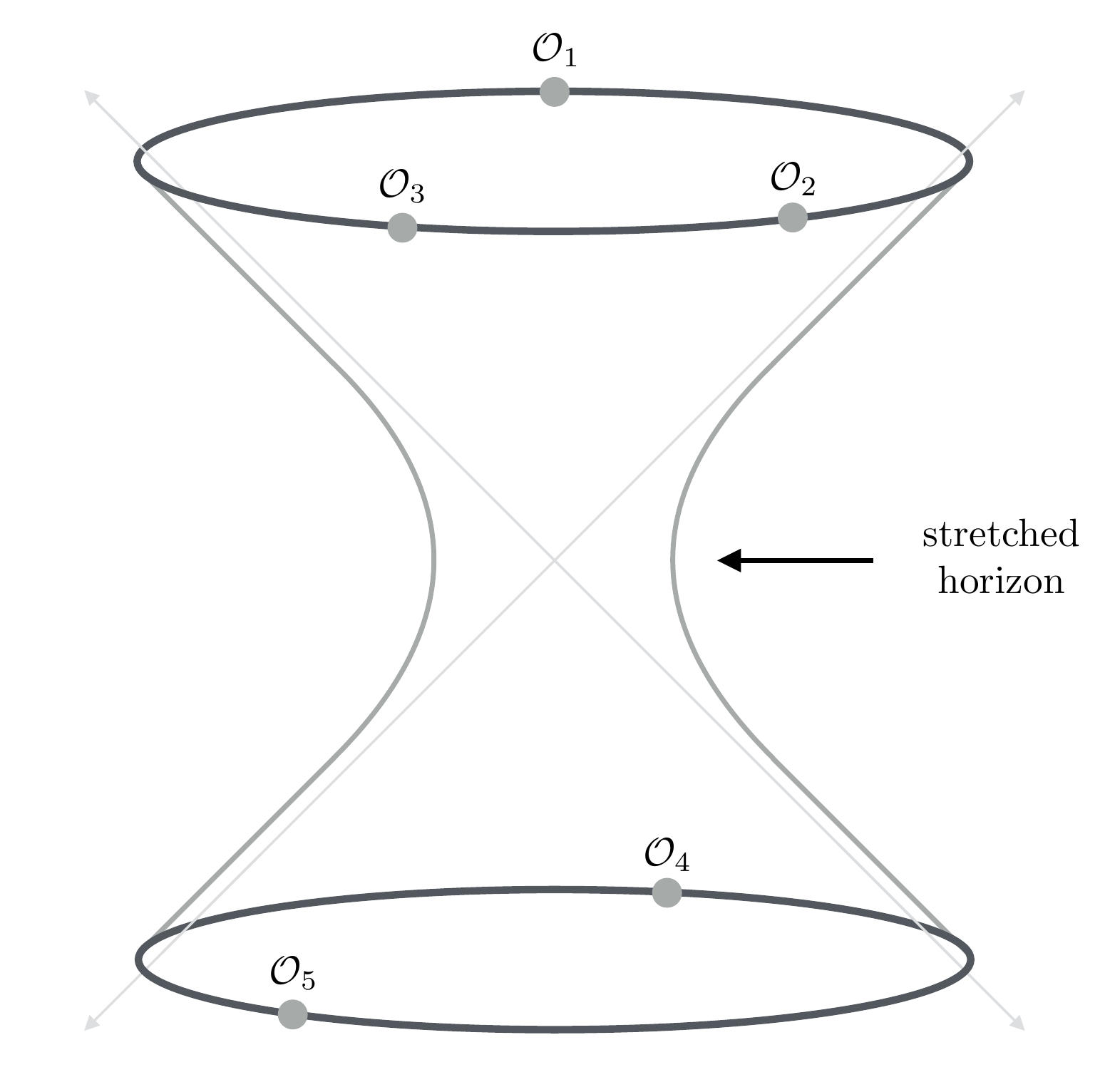}
\end{center}
\vspace*{-0.5cm}
\caption{The stretched horizon in Rindler spacetime.  The dots denote operator insertions at early and late times on the stretched horizon.}
 \label{fig:stretched_horizon}
\end{figure}

As recently discussed \cite{softHair}, it is interesting to understand in what sense asymptotic symmetries and the memory effect constitute a new kind of ``hair'' in the presence of black hole horizons. While this paper has focused on uncovering a $\CFT_2$ structure underlying $\Mink_4$ scattering amplitudes, our strategy incidentally offers a baby version of the black hole problem in the form of the Rindler horizon, say as seen by radially accelerating observers in the Rindler region. For such observers we can excise all of $\Mink_4$ spacetime that lies behind a ``stretched'' Rindler horizon, excluding the Milne regions altogether, as depicted in \Fig{fig:stretched_horizon}.

The physical observables relevant to the remaining Rindler region are {\it thermal} correlators\footnote{Here ``thermal'' is with respect to $\dS_3$ time in static patch coordinates, as experienced by a Rindler observer.} which encode the wavefunction describing the particles emitted to or from null infinity together with the stretched horizon.   First, let us remind the reader of the  $\Rind_4$ coordinates in  \Eq{eq:Rindler_metric}, where each hyperbolic slice at fixed Rindler radius $\rho$ defines a $\dS_3$ spacetime labeled by conformal time $\tau$.  As discussed earlier, the roles of $\rho$ and $\tau$ in the Rindler region are swapped relative to the Milne region.  So for any $\dS_3$ slice, the corresponding $\ddS_3$ boundary is defined by the end of time limit $\tau \rightarrow 0$.  Meanwhile, $\rho\rightarrow +\infty$ corresponds to null infinity, while $\rho = \rho_{\rm stretch}$ for large and negative $\rho_{\rm stretch}$ defines the stretched horizon.

Therefore a correlator in $\Rind_4$ has the form of a $\Mink_4$ correlator, $ \langle {\rm out} | {\cal O}_1 \cdots {\cal O}_n | {\rm in} \rangle$.  Here the in and out states label particles emitted from and to null infinity in the far past or future, respectively.    Meanwhile, the operators ${\cal O}_i$ denote insertions of particle fields on the stretched horizon at early or late times.  These operators are generic probes of the wavefunction of the stretched horizon. Despite the fact that we have restricted physical spacetime to the Rindler region outside the stretched horizon, we must compute this correlator using Minkowski Feynman diagrams in order to match the {\it thermal}  Rindler correlators.  

Such diagrams will now consist of four ingredients: the three already discussed---interaction vertices, propagators, and LSZ wave packets---together with additional propagators running from the ${\cal O}_i$ inserted in the far past or future of the stretched horizon to interactions in the bulk of the Rindler region. Since the stretched horizon at fixed $\rho =\rho_{\rm stretch}$ has Lorentzian $\dS_3$ geometry, these additional propagators describe a bulk point in $\dS_3$ and a boundary point on $\ddS_3$, so they are bulk-{\it boundary} propagators from this perspective. Therefore by the close analogy with our Milne manipulations, we see that the $\Rind_4$ correlators are boundary correlators in a $\dS_3$ theory which can be reinterpreted as dual to
 $\CFT_2$ correlators. In the standard $\dS_3/\CFT_2$ picture, the $\CFT_2$ is dual to the late time wavefunction of the Universe \cite{maldacenaNonGaussianities}.  So the $\CFT_2$ describing the Rindler region is dual to the late time
 wavefunction of Rindler, up to and including the stretched horizon and given initial conditions for the wavefunction at early times.
 
 In this context, let us analyze the physics of the CS gauge field and electromagnetic memory in the Rindler region. By the exact analog of \Eq{eq:CSfield}, we can locate the CS field in $\Rind_4$ by taking the limit of soft Rindler momentum, so the CS field corresponding to $(+)$ helicity is
 \begin{equation}
A_i(y) = A^{+}_i(\rho \rightarrow + \infty,y) - A^{+}_i(\rho =\rho_{\rm stretch},y).
\label{eq:Adiff}
 \end{equation}
The second term represents the component of the soft ``memory'' field that remembers the hard charges that fall into the Rindler horizon. We see this explicitly because, retaining this component, the analog of our AB phase associated with a region $R$ on null infinity of the Milne region in \Eq{eq:wardAdS}, now reads in  Rindler as
\eq{
\oint_{\partial R} dz \,  A_z \text(\tau\to0,  z)   =\sum_{i \in R} q_i,
 }
where again the above expression is implicitly evaluated within a 2D correlator.  Here $A_z$ is given by the two terms in \Eq{eq:Adiff}.
We thereby conclude that the AB phase measures hard charges passing through an angular region $R$, regardless of whether those hard charges are falling into the horizon or are headed out to null infinity.  If one measures the charges heading out to null infinity, the CS field will encode information on where exactly the hard charges entered the horizon. This in some sense offers a sharper form of ``hair'' \cite{softHair} compared to the usual asymptotic electric field of a black hole, which remembers the charge that has fallen into the horizon but without regard to the angle of entry.

\section{Gravity}

\label{sec:gravity}

We have described the emergence of CFT structure in gauge theory amplitudes, but of course the hallmark of a true CFT is a 2D stress tensor. 
The Sugawara construction yields a stress tensor constructed from the 2D holomorphic currents dual to soft gauge fields, but this can only be a component of the full stress tensor since it does not account for hard particle dynamics. As usual in AdS/CFT, to find the full stress tensor we must consider gravity, to which we now turn. Our aim will be to reframe many of the important aspects of 4D gravity in terms of the language of 2D CFT.  

We will follow the same basic strategy for gravity as for scalar and gauge theory, moving briskly through those aspects which are closely analogous and focusing on those which introduce major new considerations. The most important such consideration is that gravity in asymptotically flat space is not Weyl invariant, since the 4D Einstein-Hilbert action,
\eq{
S_{\rm EH} &= \frac{\mPl^2}{2} \int_{\Mink_4} d^4x \, \sqrt{-g} \, R,
\label{eq:EH_action}
}
depends on the dimensionful Planck mass, $\mPl$.  For the sake of exposition, we will often restrict to the Milne region for explicit calculations, bearing in mind that we can straightforwardly continue into the Rindler region and thus all of Minkowski space via the embedding formalism.

In any case, while the dynamics cannot be mapped into a factorizable geometry like $\AdS_3 \times \R_{\tau}$, this is merely a technical inconvenience.
As in gauge theory, one can nevertheless apply a decomposition into $\AdS_3$ and $\dS_3$ modes, resulting in 3D Witten diagrams equivalent to 4D scattering amplitudes with a particular prescription for LSZ reduction onto bulk-boundary propagators.

\subsection{Stress Tensor of $\CFT_2$}

In this section, we derive a 2D stress tensor corresponding to soft gravitons in 4D.  We will show that the Ward identity for the 2D stress tensor is a particular angular convolution of the {\it subleading} soft factor for graviton emission \cite{cachazoStrominger}.  Notably, the subleading soft fact differs from the leading factor in that it depends on the angular momentum of each external leg rather than the momentum. 

The pursuit of a 2D stress tensor will naturally lead us to the Virasoro algebra, which directly manifests the super-rotation \cite{belgiansSuperrotations} asymptotic symmetries of 4D Minkowski space. Commuting these with ordinary translations, we then derive the BMS super-translations \cite{bms1,bms2}. This approach is anti-historical, but more natural from the holographic approach taken here.

\label{sec:stresstensor}

\subsubsection{Bulk-Boundary Propagator for $\AdS_3$ Graviton} 

By the $\AdS_3/\CFT_2$ dictionary, the 2D stress tensor is a tensor primary operator of scaling dimension $\Delta=2$  dual to a massless tensor field, $h_{ij}$, in $\AdS_3$.  Again using the embedding formalism, we write down the bulk-boundary propagator for $h_{ij}$ lifted from 3D to 4D via
\eq{
K_{\mu\nu} =   N(x^2) \times   \frac{x^\rho x^\sigma f_{\rho \mu} f_{\sigma \nu}}{(kx)^{4}}.
}
Here, the normalization $N(x)$ parameterizes an inherent ambiguity in the lift, arising because $\AdS_3$ lives on the constrained surface $x^2 = -1$. For gauge theory we sidestepped this ambiguity, since the underlying Weyl invariance implied that the dynamics are independent of the scale set by the constrained surface.  However, there is no such invariance of 4D gravity due to the dimensionful gravitational constant, so we must find an alternate way to identify $N(x)$. 

Of course, $N(x)$ should be chosen so $K_{\mu\nu}$ is a solution of the linearized Einstein's equations in $\Mink_4$.  Imposing this condition leaves two possibilities: either $N(x) =1$ or $N(x) = x^2$.  A priori, either solution is reasonable, but as will see, the latter is the correct choice.  The reason for this is that in standard $\AdS_3$ gravity the Virasoro symmetries arise as asymptotic symmetries of $\AdS_3$ encoded in solutions to the 3D Einstein's equations. Famously, all such solutions are pure gauge \cite{3dGravIsTrivial1,3dGravIsTrivial2}, and are thus diffeomorphisms of $\AdS_3$ itself. At the linearized level this is reflected in the fact that the bulk-boundary propagator for $\Delta=2$ on $\AdS_3$ is a linearized 3D diffeomorphism about AdS$_3$.  In order to recast the Virasoro symmetries as asymptotic symmetries of $\Mink_4$, we should look for a lift of the $\AdS_3$ bulk-boundary propagator that yields a pure linearized large diffeomorphism in $\Mink_4$. 

A straightforward calculation shows that for the bulk-boundary propagator $K_{\mu\nu}$ is not a pure 4D diffeomorphism for $N(x)=1$, but is for $N(x)=x^2$.   This is reasonable since the $\Milne_4$ is a warped product of $\AdS_3$ and $\R_\tau$ associated with a warp factor $x^2 = e^{2\tau}$, which we now see is crucial to lift 3D diffeomorphisms into a 4D diffeomorphisms.   Fixing $N(x) = x^2$, our final expression for the lifted bulk-boundary propagator for $\Delta=2$ is
\eq{
K_{\mu\nu} =  x^2 \times \frac{x^\rho x^\sigma f_{\rho \mu}  f_{\sigma \nu}}{(kx)^{4}}.
\label{eq:K_grav}
}
Since this is a pure diffeomorphism, it can be written as
\eq{
K_{\mu\nu} =  \partial_\mu \xi_\nu+\partial_\nu \xi_\mu   \qquad \textrm{where} \qquad \xi_\mu =\frac{1}{3} \partial^3_z (x^\rho\bar f_{\rho\mu}  \log kx),
\label{eq:K2def}
}
where $\bar f_{\mu\nu}$ is defined in \Eq{eq:fdefs}.  This form for $K_{\mu\nu}$ will be quite useful for explicit calculations. 

Applying the logic of AdS/CFT, the bulk-boundary propagator for $h_{ij}$ corresponds to the insertion of a local $\CFT_2$ stress tensor $t(z)$ or its complex conjugate $\bar t(\bar z)$. In the subsequent sections, we will see  how the bulk-boundary propagator $K_{\mu\nu}$ relates to single and multiple soft graviton emission in 4D. 

Finally, let us comment on the curious fact the bulk-boundary propagator for gravity is proportional to the square of the bulk-boundary propagator for gauge theory, so
\eq{
K_{\mu\nu} = x^2 K_\mu   K_\nu.
}
The simplicity of this is remarkable, given the known (gauge)$^2$ =  gravity relations that arise from the KLT \cite{KLT} relations and the closely related BCJ \cite{BCJ} relations.  Given also the connection between BCJ and the soft limit \cite{WhiteBCJ}, it is likely that the above equation is not an accident, and is perhaps a sign of some deeper underlying construction.

\subsubsection{Ward Identity for $\CFT_2$ Stress Tensor}

Given the central role of the 2D stress tensor $t(z)$, it is natural to ask about the 4D dual of this quantity. Repeating our strategy for gauge theory, we now calculate the Ward identity for the 2D stress tensor using AdS/CFT.  To do so, we compute a correlator of the stress tensor via the associated Witten diagram,
\eq{
\langle t(z) {\cal O}(z_1,\bar z_1)\cdots {\cal O}(z_n,\bar z_n)\rangle = \int d^4 x \, K_{\mu\nu}(x)W^{\mu\nu}.
}
Here $K_{\mu\nu}$ is the bulk-boundary propagator in \Eq{eq:K_grav} and $W^{\mu\nu}$ parameterizes the remainder of the Witten diagram,
\eq{
W^{\mu\nu} = \langle \textrm{out} |T^{\mu\nu}(x) | \textrm{in} \rangle,
}
computed as an insertion of the 4D stress tensor operator $T^{\mu\nu}$ inserted between in and out states.

Substituting the pure gauge form of the bulk-boundary propagator in \Eq{eq:K2def}, we obtain
\eq{
\langle t(z)  {\cal O}(z_1,\bar z_1)\cdots {\cal O}(z_n,\bar z_n) \rangle &=  \int d^4 x\, [\partial_\mu\xi_\nu(x)+\partial_\nu\xi_\mu(x) ]  \langle \textrm{out} |T^{\mu\nu}(x) | \textrm{in} \rangle \nonumber \\& \hspace*{-1cm} =- \frac{1}{3}\partial^3_z \left(  \int d^4 x\,   \bar f_{\mu \nu}  \log kx \, \partial_\rho  \langle \textrm{out} |x^\mu T^{\rho\nu}(x)-x^\nu T^{\rho \mu}(x)| \textrm{in} \rangle \right),
}
where in the second line we have shuffled around terms and performed an integration by parts, dropping boundary terms.  Importantly, the expression sandwiched between in and out states is the relativistic angular momentum tensor.  This quantity is conserved everywhere except at insertions associated with the external legs, so
\eq{
\partial_\rho  \langle \textrm{out} |x^\mu T^{\rho\nu}(x)-x^\nu T^{\rho \mu}(x)| \textrm{in} \rangle  = -\sum_{i=1}^n J^{\mu\nu}_i \delta^4(x-x_i),
\label{eq:ang_nonconservation}
}
where $J^{\mu\nu}_i$ is the angular momentum of each external particle and $x_i$ is its insertion point near the boundary. As before, we substitute the position of the external particles inserted near the boundary $\rho_i \rightarrow 0$ with their corresponding momenta, so $x_i \sim k_i$.  As a result, the expression for the Ward identity will involve manifestly on-shell quantities.  Plugging this substitution into the Ward identity, we obtain
\eq{
\langle t(z)  {\cal O}(z_1,\bar z_1)\cdots {\cal O}(z_n,\bar z_n)\rangle =\frac{1}{3}  \partial^3_z   \left( \sum_{i=1}^n    \log kk_i  \, \bar f_{\mu \nu} J^{\mu\nu}_i\right) \langle {\cal O}(z_1,\bar z_1)\cdots {\cal O}(z_n,\bar z_n)\rangle .
\label{eq:t_simp}
}
In the above equation, the angular momentum generator is implicitly defined in momentum basis, so {\it e.g.} it acts on a hard scalar leg as 
\eq{
J_i^{\mu\nu} = k_i^{\mu} \frac{\partial}{\partial k_{i\nu}}-k_i^{\nu} \frac{\partial}{\partial k_{i\mu}}.
\label{eq:Jinmom}
} 
The analogous expression for hard legs with spin has a simple representation in terms of spinor helicity variables. 
From \Eq{eq:t_simp} we see directly the connection between the stress tensor in the $\CFT_2$ and rotations acting on the boundary of $\Mink_4$.  This is not accidental, and as we will see later is a hint of the super-rotation asymptotic symmetries of 4D flat space.

To compare this to the usual 2D stress tensor Ward identity, it is actually convenient to briefly revert to position space for the hard particles.  To do so we send $k_i \sim x_i$ in \Eq{eq:t_simp} and \Eq{eq:Jinmom} and go to Milne coordinates.  Taking the $\rho_i\rightarrow 0$ limit, we find
\eq{
\langle t(z)  {\cal O}(z_1,\bar z_1)\cdots {\cal O}(z_n,\bar z_n) \rangle \sim  \sum_{i=1}^n \left[  \frac{h_i}{(z-z_i)^2} + \frac{1}{z-z_i}  \frac{\partial}{\partial z_i}  \right] \langle {\cal O}(z_1,\bar z_1)\cdots {\cal O}(z_n,\bar z_n)\rangle,
\label{eq:t_ward}
} 
where the conformal weight is
\eq{
h_i = \left. \frac{\partial}{\partial \log \rho_i^2} \right\vert_{\rho_i\rightarrow 0},
}
for a 2D scalar operator dual to a hard 4D scalar particle.
Up to an overall constant normalization, \Eq{eq:t_ward} is none other than the Ward identity for the stress tensor of the $\CFT_2$.  
  Of course, this analysis can be extended straightforwardly to include hard particles with spin.

\subsubsection{Relationship to Subleading Soft Theorems in $\Mink_4$}

Next, we derive the explicit relationship between the Ward identity for the 2D stress tensor and the soft graviton theorems.  To do so, it will be convenient introduce an auxiliary operator $\tilde t(z,\bar z)$ which is a $\Delta=0$ tensor primary operator of the 2D CFT.  Note that we do not assign {\it independent} physical import to this $\Delta =0$ operator, which is why we refer to it as auxiliary. 
 
From the embedding formalism, the bulk-boundary propagator for $\tilde t(z,\bar z)$ is
\eq{ 
\tilde K_{\mu\nu} =  \frac{x^\rho x^\sigma f_{\rho \mu}  f_{\sigma \nu}}{(kx)^2} .
\label{eq:tildeK}
}
Importantly, this bulk-boundary propagator is a pure linearized diffeomorphism equal to
\eq{
\tilde K_{\mu\nu} = \partial_\mu \tilde\xi_\nu+\partial_\nu \tilde\xi_\mu \qquad \textrm{where} \qquad 
\tilde \xi_\mu = \frac{1}{2} \partial_z(x^\rho f_{\rho\mu} \log kx).
\label{eq:tildeK2def}
}
Repeating our steps from before, calculate an arbitrary correlator involving $\tilde t(z,\bar z)$,
\eq{
\langle \tilde t(z, \bar{z})  {\cal O}(z_1,\bar z_1)\cdots {\cal O}(z_n,\bar z_n) \rangle &\sim  \partial_z   \left( \sum_{i=1}^n    \log k k_i  \, f_{\mu \nu} J^{\mu\nu}_i\right) \langle {\cal O}(z_1,\bar z_1)\cdots {\cal O}(z_n,\bar z_n)\rangle  \nonumber \\
& =  \sum_{i=1}^n \frac{\epsilon k_i}{k k_i}f_{\mu \nu} J^{\mu\nu}_i \langle {\cal O}(z_1,\bar z_1)\cdots {\cal O}(z_n,\bar z_n)\rangle,
\label{eq:tilde_t}
}
where the right-hand side is literally the subleading graviton soft factor \cite{cachazoStrominger}.   While interesting, this observation is only useful because $\tilde t(z,\bar z)$ happens to be directly related to $t(z)$ by a handy integral transform in $(z,\bar z)$.  Indeed,
by comparing the definitions of $\xi_\mu$ and $\tilde \xi_\mu$ in \Eq{eq:K2def} and \Eq{eq:tildeK2def}, respectively, we see that these quantities are related by the differential equation,
\eq{
\partial_{\bar z} \xi_\mu \sim \partial^3_z \tilde \xi_\mu^*,
\label{eq:xi_diffeq}
}
dropping unimportant numerical prefactors.  Notably, the above equation is equivalent to the $\CFT_2$ equation $\partial_{\bar z} t(z) \sim \partial^3_z \tilde t^\dagger(z,\bar z)$, which when evaluated inside a correlator yields zero on both sides except for delta function support at the insertion points of hard operators.  In fact, we can verify this fact by applying $\partial^3_{\bar z}$ directly to \Eq{eq:tilde_t}.  Since $\partial^3_{\bar z} f_{\mu\nu} = 0$, this implies that at least one $\partial_{\bar z}$ derivative will act on $\epsilon k_i / k k_i = 1/(z-z_i)$, producing a delta function $\delta^2(z-z_i)$ from the identity in \Eq{eq:d1z}.  This is a non-trivial check that the structure of the subleading graviton soft theorem ensures conservation of the $\CFT_2$ stress tensor.

In any case, we would like to solve the differential equation in \Eq{eq:xi_diffeq} by constructing a formal anti-derivative,
\eq{
\partial^{-1}_{\bar z} = \frac{1}{2\pi}\int d^2 z' \frac{1}{z-z'},
}
which satisfies $ \partial_{\bar z} \partial^{-1}_{\bar z}=1$ as a result of \Eq{eq:d1z}.
Solving \Eq{eq:xi_diffeq} then yields
\eq{
\xi_\mu(z,\bar z) \sim  \partial_z^3 \int d^2 z' \frac{1}{z-z'}\, \tilde\xi^*_\mu(z', \bar z'),
}
suppressing all $\tau$ and $\rho$ dependence.  Inserting this relation into the Ward identity for the stress tensor, we obtain our final expression,
\eq{
\langle  t(z) {\cal O}(z_1,\bar z_1)\cdots {\cal O}(z_n,\bar z_n) \rangle &\sim \partial^3_z \int d^2 z' \frac{1}{z-z'}\, \langle \tilde{t}(z, \bar{z})^{\dagger}  {\cal O}(z_1,\bar z_1)\cdots {\cal O}(z_n,\bar z_n) \rangle \nonumber \\  
& \hspace*{-1cm} \sim   \partial^3_z \int d^2 z' \frac{1}{z-z'}\, \sum_{i=1}^n   \frac{\bar \epsilon' k_i}{k' k_i}\bar f_{\mu \nu}' J^{\mu\nu}_i \langle  {\cal O}(z_1,\bar z_1)\cdots {\cal O}(z_n,\bar z_n)\rangle,
\label{eq:t_to_tildet}
}
where $k'$, $\bar \epsilon'$, and $\bar f'$ are functions of $(z', \bar z')$.  This result says that the Ward identity for the 2D stress tensor is proportional to a particular angular integral over the subleading soft graviton factor.  Physically, this corresponds to a particular superposition of soft graviton emission in all directions $(z', \bar z')$.  

Let us pause to discuss the peculiar integral structure of \Eq{eq:t_to_tildet}.  Naively, it is odd that the $\CFT_2$ stress tensor should be expressed as a non-local function in $(z,\bar z)$ but this was actually essential to maintain consistency between the 2D and 4D pictures.  To see why, recall from \Eq{eq:t_ward} that the canonical form of the 2D stress tensor Ward identity has manifest double and single poles in $z$.  In turn, this OPE corresponds to collinear singularities in 4D, but graviton scattering amplitudes are famously free of such collinear singularities.  Hence, the only way to square these apparently inconsistent statements is if the 2D stress tensor is actually a {\it non-local} function of the graviton scattering amplitude in $(z,\bar z)$, as \Eq{eq:t_to_tildet} clearly is.  Only then is it possible for the singularity structure of the 2D stress tensor Ward identity to arise consistently from the analytic properties of graviton amplitudes.

\subsection{Virasoro Algebra of $\CFT_2$}
\label{sec:virasoro}

The Virasoro algebra places immense constraints on the structure of correlators in the $\CFT_2$.  It is obviously of great interest to understand the implications of these constraints on the dual scattering amplitudes in $\Mink_4$.   As we will see, the corresponding infinite-dimensional Virasoro algebra in 2D has a direct connection to the asymptotic symmetries of 4D flat space \cite{belgiansSuperrotations,stromingerVirasoro}.

What is the action of the Virasoro generators on scattering amplitudes?  To answer this, we revisit the 2D stress tensor Ward identity in \Eq{eq:t_simp}.  Expanding the derivatives in $z$, we obtain
\eq{
\langle t(z)  {\cal O}(z_1,\bar z_1)\cdots {\cal O}(z_n,\bar z_n) \rangle = \nonumber\\
& \hspace*{-3.5cm} \sum_{i=1}^n  \left( \frac{2}{3} \left(\frac{\epsilon k_i}{kk_i} \right)^3 -  \left(\frac{\epsilon k_i}{kk_i} \right)^2 \partial_z  +  \left(\frac{\epsilon k_i}{kk_i} \right)\partial_z^2 \right) \bar f_{\mu\nu} J^{\mu\nu}_i \langle {\cal O}(z_1,\bar z_1)\cdots {\cal O}(z_n,\bar z_n)\rangle.
}
For simplicity, consider the limit in which the soft graviton is collinear to a hard external leg located at $z'$ on the celestial sphere.  A Laurent expansion of this expression around $z= z'$ yields
\eq{
\langle t(z)  {\cal O}(z',\bar z') \cdots \rangle =  \left( \frac{2 \bar f_{\mu\nu}'/3}{(z-z')^3}  - \frac{\partial_{z'} \bar f_{\mu\nu}'/3}{(z-z')^2} + \frac{\partial^2_{ z'} \bar f_{\mu\nu}'/3}{z-z'} + \ldots \right) J^{\prime \mu\nu} \langle {\cal O}(z',\bar z')\cdots \rangle ,
\label{eq:Laurent_t}
}
where all primed quantities are evaluated at $z=z'$ and we have used \Eq{eq:exkx}.  Here the ellipses denote non-singular contributions which originate from the other hard legs in the process.

We can now compare \Eq{eq:Laurent_t} directly to definition of the Virasoro generators, 
\eq{
t(z) = \sum_{m=-\infty}^\infty  \frac{L_m}{z^{m+2}} = \ldots + \frac{L_{1}}{z^3}  +\frac{L_{0}}{z^2} +\frac{L_{-1}}{z}  +\ldots,
}
only Laurent expanded around $z=z'$. Matching terms by eye, we ascertain the identities of the $SL(2,\mathbb{C})$ Virasoro generators, 
\eq{
L_1 &\sim -i(K_2 + i J_2)-(K_1 + i J_1) \nonumber \\
L_0 &\sim K_3 + i J_3 \nonumber \\
L_{-1} &\sim -i(K_2 + i J_2)+ (K_1 + i J_1),
}
up to a constant normalization factor.  Here $K_3$ and $J_3$ denote the generators of  $J^{\prime \mu\nu}$ corresponding to boosts and rotations around the axis of the hard particle, while $K_{1,2}$ and $J_{1,2}$ are those for the transverse directions.  Since these generators only act on the collinear hard particle, they are effectively local Lorentz transformations.  Thus, the identification of the full Virasoro algebra as the algebra of super-rotations is indeed appropriate.

This result offers a physical interpretation for the action of $t(z)$ on scattering amplitudes.   The passage of collinear emitted soft gravitons induces a Lorentz transformation that acts locally on a hard leg.  Operationally, this ``jiggles'' the hard particle in a way that displaces it relative to the direction of its original trajectory.  This local Lorentz transformation has the same effect as  a net displacement of the detectors residing at the boundary of spacetime.

\subsection{Chern-Simons Theory and Multiple Soft Emission} 

\label{sec:gravityCS}

To understand multiple soft emissions in gravity, we proceed in parallel with our analysis for non-abelian gauge theory.  Our aim is to describe the dynamics of multiple external soft gravitons that interact and merge in the gravitational analog of \Fig{fig:soft_emissions}.
As before, we can parameterize the dynamics of the entire soft branch with a graviton field $H_{\mu\nu}(x)$ at the juncture $x$ with the hard process characterized by $T_{\mu\nu}(x)$.  In the limit of vanishing gravitational coupling, $H_{\mu\nu}$ will approach a superposition of independent soft gravitons, each described by the bulk-boundary propagator $K_{\mu\nu}$ from \Eq{eq:K_grav}.   Hence, the branch structure of soft gravitons is rooted in external legs connected through these bulk-boundary propagators.  Said another way, the soft branch is simply the solution to the {\it non-linear} sourceless Einstein's equations with free-field approximation given by $K_{\mu \nu}$.  

Now consider a closely analogous situation for 3D Witten diagrams, where an $\AdS_3$ branch field $h_{ij}(y)$ similarly characterizes the web of soft gravitons merging before making contact with a hard source at $y$.  Here $h_{ij}$ can be treated as a perturbation of the background $\AdS_3$ metric $g_{ij}$ defined in \Eq{eq:hyper_coords}.  The full metric in 3D is then
\eq{
\tilde g_{ij} = g_{ij} + h_{ij},
\label{eq:backgroundgrav}
} 
where $g_{ij}$ is the background $\AdS_3$ metric from \Eq{eq:hyper_coords}.  \Eq{eq:backgroundgrav}
is a solution to Einstein's equations in $\AdS_3$ whose free field asymptotics near $\dAdS_3$ are given by bulk-boundary propagators.  Since all solutions to $\AdS_3$ gravity are pure diffeomorphisms of $\AdS_3$ \cite{3dGravIsTrivial1,3dGravIsTrivial2}, $h_{ij}$ corresponds to precisely such a non-linear diffeomorphism.  

Next, using the same prescription as for bulk-boundary propagators, we can lift this diffeomorphism from 3D to 4D.  In particular, we have that $H_{ij} = e^{2\tau} h_{ij}$, where $H_{ij}$ are the non-zero components of the 4D branch field $H_{IJ}$ in Milne temporal gauge.   The $x^2 = e^{2\tau}$ warp factor is the same one required in the bulk-boundary propagator for the 2D stress tensor.  Since $H_{IJ}$ is a 4D diffeomorphism around flat space, we find 
\begin{equation} 
(\eta_{\mu\nu}+ H_{\mu\nu}(x)) d x^{\mu} d x^{\nu}  =   e^{2 \tau} ( -d \tau^2 + \tilde g_{ij}(y) dy^i dy^j).
\label{eq:minkDiff}
\end{equation}
In conclusion, at the fully non-linear level, multiple subleading soft emissions are described by a branch $H_{\mu\nu}$ that encodes large diffeomorphisms of the $\AdS_3$ metric.  

Since these soft perturbations of the metric are Milne zero modes, they couple to hard particle tracks according to
\eq{
\int d^4 x \,H_{\mu \nu}(x) T^{\mu \nu}(x) = \int d^3 y \, \sqrt{g} \,  {h}_{ij}(y) \int d \tau\, e^{6 \tau} T^{ij}(\tau, y) = \int d^3 y \, \sqrt{g} \,  {h}_{ij}(y) T^{ij}_{\rm eff}(y),
}
where in the last line we have defined 
\begin{equation}
T^{ij}_{\rm eff}(y) =  \int d \tau \, e^{6 \tau} T^{ij}(\tau, y) ,
\end{equation}
the Milne time-integrated stress tensor in a warped version of \Eq{eq:K_grav}.

\subsubsection{Equivalence to $\AdS_3$ Gravity}

Similar to the case of gauge theory, we have seen that 4D soft graviton modes correspond to solutions of 3D gravity which are pure diffeomorphisms.  It is then expected that the resulting theory is topological, which is reasonable because gravity in $\AdS_3$ is famously equivalent to a CS theory, at least perturbatively \cite{witten3dGravity1}.  In particular, one can define a non-abelian CS gauge field, $A_i^{ \pm a} = \epsilon^{abc} \Omega_{ibc} \pm i e^a_i$, where $e$ is the dreibein, $\Omega$ is the spin connection, and  the index $a= 1,2,3$ runs through the local tangent space. The gauge group of the CS theory is $SL(2, \mathbb C)$, corresponding to the global isometries of $\AdS_3$, or equivalently, the Lorentz group in $\Mink_4$.  Concretely, $A_i^{\pm a}$ corresponds to the Lorentz generators $J^a \pm i K^a$, where $J^a$ and $K^a$ are rotations and boosts, respectively. 

Via the embedding formalism, 
$A_i^{a \pm}$ is associated with $(+)$ and $(-)$ helicity soft gravitons.   Moreover, since the commutator $[J^a + i K^a, J^b - i K^b] = 0$ vanishes, the $SL(2, \mathbb C)$ gauge group factorizes, so there is no intrinsic reason why we must restrict to a single helicity like we did for non-abelian CS theory.  
At the level of the dual $\CFT_2$ we are then permitted to compute mixed correlators involving both the holomorphic and anti-holomorphic stress tensor, $t(z)$ and  $\bar{t}(\bar{z})$.

With the non-abelian structure clarified, we can Laurent expand the holomorphic stress tensor into the infinite set of non-abelian Virasoro charges. 
Relatedly, the CS structure of the subleading soft amplitudes again implies that the dynamics of soft gravitons is governed by a non-abelian analog of the AB effect, where the CS graviton field is the now the field  encoding memory effects. Unlike for electromagnetic memories, we have not as yet matched this kind of AB effect in detail with the ``spin memory'' effects discussed already in the literature \cite{stromingerVirasoroMemory}.

A final note on the rigor of our conclusions here: what we have shown thus far is that LSZ reduction onto $\Delta=2$ bulk-boundary propagators  gives a consistent picture for multiple subleading soft emissions. We have not yet proven that the scattering amplitudes of plane waves have the requisite commutativity amongst multiple subleading soft limits required for simultaneous LSZ reduction onto multiple bulk-boundary propagators. But we expect that the $\AdS_3$ gravity picture should identify any obstructions to multiple soft limits, as it did in non-abelian CS gauge theory for mixed soft helicities. While no such obstructions appear here, it would still be interesting to compute explicitly the commutativity properties of subleading graviton soft limits for these amplitudes in Minkowski space.

\subsubsection{Virasoro Central Charge from Internal Soft Exchange}
 
The Virasoro central charge, $c$, is arguably the most important quantity in a 2D CFT \cite{cardyC}.   In theories with semi-classical $\AdS_3$ duals, $c$ is given by the $\AdS_3$ Planck scale in units of the $\AdS_3$ length.  However, much like the gauge coupling in YM theory, the Planck scale enters simply as an overall factor in the gravity action, so it drops out of the homogeneous Einstein's equations.  So at tree level, the soft branches characterizing multiple graviton emission are insensitive to the Planck scale and thus $c$. 

To make sense of $c$, we must then consider the gravitational analog of \Fig{fig:soft_exchange}, which depicts a set of two hard processes exchanging a soft internal graviton.  We interpret one process as a ``measurement apparatus'' for the subleading soft graviton emission of the other. Notably, the corresponding $\AdS_3$ Witten diagram is suppressed by $1/c$, while the $\Mink_4$ scattering amplitude goes as $1/\mPl^2$. However, unlike before when we matched the CS level to the gauge coupling, here there is a dimensional mismatch between $c$ and $\mPl^2$.
This means that an infrared length scale $L_{\rm IR}$ does not decouple from the process.  One can think of $L_{\rm IR}$ as a formal scale separating ``hard'' from ``soft''.  We thereby conclude that the Virasoro central charge scales as
\begin{equation} \label{eq:ceff}
c \sim \mPl^2 L_{\rm IR}^2.
\end{equation}
Just this type of infrared sensitivity is present in the spin-memory effect described in \cite{stromingerVirasoroMemory}.

We can see this more directly by writing the 4D Einstein-Hilbert action in \Eq{eq:EH_action} in terms of the 3D metric $\tilde g_{ij}$ characterizing a soft branch in $\AdS_3$, as shown in \Eq{eq:backgroundgrav}.   Since $\tilde g_{ij}$ is related by a diffeomorphism to the pure $\AdS_3$ background metric, the resulting action should just be proportional to 3D gravity with a cosmological constant.   The simple $\tau$ dependence of the action straightforwardly factors, yielding
\eq{
S_{\rm EH} &= \frac{\mPl^2}{2} \int_{\Milne_4} d^4Y \, \sqrt{- G} \,  R = \frac{\mPl^2}{2} \int_{- \infty}^{\tau_{\rm late}} d \tau \, e^{2 \tau} \int_{\AdS_3} d^3y \, \sqrt{\tilde{g}}  \, (\tilde{R} + 1), 
}
where we have taken ``unit'' dimensionally reduced $\AdS_3$ radius of curvature, in keeping with the normalization of our other formulas, and 
where $\tau_{\rm late}$ relates to $L_{\rm IR}$ by
\eq{
L_{\rm IR} \sim  e^{\tau_{\rm late}}.
}
Although we are not carefully treating the physics underlying $\tau_{\rm late}$ here, we can nevertheless estimate the central charge from this rough scaling,
\begin{equation}
c\sim \frac{\mPl^2}{2} \int_{- \infty}^{\tau_{\rm late}} d \tau \, e^{2 \tau} \sim \mPl^2 e^{2\tau_{\rm late}} \sim  \mPl^2 L_{\rm IR}^2.
\end{equation} 
We leave a formal analysis of the Virasoro central charge for future work.

\subsection{Relation to Asymptotic Symmetries}

\subsubsection{From Super-Rotations to Super-Translations in $\Mink_4$}

Let us now discuss the relation between our results and the asymptotic symmetries of $\Mink_4$.  While there is an expansive literature on this subject, we will be quite brief here.   Long ago, BMS \cite{bms1,bms2}
discovered the existence of an infinite-dimensional symmetry of asymptotically flat space corresponding to super-translations at null infinity.  Physically, these super-translations are diffeomorphisms of retarded time that depend on angles on the celestial sphere.

More recently, \cite{belgiansSuperrotations} argued that the super-translation algebra can be further extended to include super-rotations encoding an underlying Virasoro algebra.  From their analysis of large diffeomorphisms, they proposed an extended BMS algebra \cite{belgiansChargeAlgebra}, 
\eq{
[L_m,L_n]&=(m-n)L_{m+n} \nonumber \\
[P_{mn},P_{rs}]&=0  \nonumber\\
[L_m,P_{rs}]&= \left(\frac{m+1}{2} -r\right) P_{m+r,s},
\label{eq:algebra}
}
dropping for the moment the Virasoro central charge.  
Here the Virasoro generators $L_m$ correspond to the super-rotations while the generators $P_{mn}$ correspond to super-translations.  The Poincare sub-algebra is
\eq{
L_{-1}, L_{0}, L_{1},\qquad \bar L_{-1}, \bar L_{0}, \bar L_{1},\qquad P_{00},P_{01},P_{10},P_{11},
}
where the four super-translation generators are nothing more than the four components of the usual momentum generator $P_{\alpha \dot \alpha}$ in the spinor basis where $\alpha,\dot\alpha=0,1$. Ref.~\cite{stromingerBMS1,stromingerBMS2,stromingerVirasoro} later showed that the super-translations and super-rotations, at least at the level of single soft emission, arise from the leading and subleading Weinberg soft theorems. 

Here we will use \Eq{eq:algebra} as a guide for constructing super-translations as a combination of super-rotations and ordinary translations.  
While ordinary translations are quite obscure in Milne and Rindler coordinates, they are of course still a symmetry of flat space, so they should also be global symmetries of the CFT.  Since the 2D stress tensor is comprised of super-rotation generators, we can commute it with regular translations to obtain
\eq{
[t(z) , P_{00}] &= \sum_{m=-\infty}^{\infty} \frac{1}{z^{m+2}} [L_m,P_{00}]
= \frac{1}{2}\sum_{m=-\infty}^{\infty} \frac{m+1}{z^{m+2}} P_{m0} =   -\frac{\partial_z j(z)}{2} .
\label{eq:tp_comm}
}
In analogy with the 2D CFT for gauge theory, we have defined a super-translation current,
\eq{
j(z) = \sum_{m=-\infty}^\infty \frac{P_{m0}}{z^{m+1}},
}
which is holomorphically conserved, so $\partial_{\bar z} j(z)=0$.

We can use this result to determine the Ward identity for $j(z)$.  From our formula for the 2D stress tensor Ward identity in \Eq{eq:t_simp}, we already see an explicit connection to super-rotations through the angular momentum operators $J_i^{\mu\nu}$ acting on the hard legs.  Now taking the commutator of \Eq{eq:t_simp} with $P_{00}$, we obtain
\eq{
\langle [t(z),P_{00}]  {\cal O}(z_1,\bar z_1)\cdots {\cal O}(z_n,\bar z_n)\rangle & \sim  \partial_z^3   \left( \sum_{i=1}^n    \log kk_i  \, \bar f_{\mu \nu}  [J_i^{\mu\nu} , k_i^\rho] q_\rho\right) \langle {\cal O}(z_1,\bar z_1)\cdots {\cal O}(z_n,\bar z_n)\rangle .
\label{eq:correlator_comm}
}
Here we have used that $P_{00} = q P$ to go from explicit  spinor index notation to a more covariant form.  We can evaluate this expression using the fact that global translations $P^\mu = \sum_{i=1}^n k^\mu_i$ have a non-vanishing commutator with the angular momentum generators acting on the hard legs,
\eq{
[J^{\mu\nu}_i , k_i^\rho] =k_i^\mu \eta^{\nu\rho}  - k_i^\nu\eta^{\mu\rho} .
}
Applying these relations, \Eq{eq:correlator_comm} simplifies to
\eq{
 &\sim  \partial_z   \left( \sum_{i=1}^n  \frac{\epsilon k_i}{k k_i} qk_i  \right) \langle {\cal O}(z_1,\bar z_1)\cdots {\cal O}(z_n,\bar z_n)\rangle = \partial_z   \left( \sum_{i=1}^n  \frac{qk_i}{z-z_i}  \right) \langle {\cal O}(z_1,\bar z_1)\cdots {\cal O}(z_n,\bar z_n)\rangle .
\label{eq:correlator_comm}
}
Comparing with \Eq{eq:tp_comm}, we see that
the Ward identity for the super-translation current is
\eq{
\langle j(z)  {\cal O}(z_1,\bar z_1)\cdots {\cal O}(z_n,\bar z_n) \rangle &\sim  \sum_{i=1}^n  \frac{qk_i}{z-z_i}  \langle {\cal O}(z_1,\bar z_1)\cdots {\cal O}(z_n,\bar z_n) \rangle .
\label{eq:STward}
}
Hence, we deduce that the charge associated with the super-translation Ward identity is the physical momentum in the $q$ direction.

\subsubsection{Chern-Simons Theory for Super-Translations?}

We have shown how 4D super-translations can be obtained from the 2D stress tensor $t(z)$ via the commutation relations of the extended BMS algebra.   Furthermore, we saw that correlators of $t(z)$ correspond to a particular angular convolution of the subleading graviton soft theorem.  Given the underlying connection of $j(z)$ to super-translations, it is then quite natural for $j(z)$ to relate to the {\it leading} graviton soft theorem.  As we will see, this is indeed the case.  
 
 To understand why, we revisit the auxiliary tensor primary $\tilde{t}(z,\bar z)$ defined in \Eq{eq:tilde_t}, whose correlators are literally equal to the 4D subleading soft graviton factor.  In particular, let us consider the $\CFT_2$ operator, $[\tilde t(z, \bar{z}), P_{00}]$, defined by the commutator of this auxiliary tensor and regular translations.  
 
It is simple to see that the bulk-boundary propagator associated with the operator $[\tilde t(z, \bar{z}), P_{00}]$ is a pure diffeomorphism.  In particular, since $P_{00}=qP$ the bulk-boundary propagator for $[\tilde t(z, \bar{z}), P_{00}]$ is by definition just the derivative of the bulk-boundary propagator of $\tilde t(z,\bar z)$ in the $q$ direction.  Concretely, this implies that the bulk-boundary propagator for $[\tilde t(z, \bar{z}), P_{00}]$  is simply $q^{\rho} \partial_{\rho} \tilde{K}_{\mu \nu}$, where $\tilde K_{\mu\nu}$ is the bulk-boundary propagator for $\tilde t(z,\bar z)$. Since the latter is a pure diffeomorphism, so too is the former.  
As we will see, this happens for a reason: this commutator is directly related to the holomorphic current for super-translations, $j(z)$.

Using our now standard methodology, let us compute the correlator for this commutator,
\eq{
\langle [\tilde t(z, \bar{z}), P_{00}]  {\cal O}(z_1,\bar z_1)\cdots {\cal O}(z_n,\bar z_n) \rangle & \sim \sum_{i=1}^n \frac{\epsilon k_i}{k k_i}f_{\mu \nu} [J_i^{\mu\nu} , k_i^\rho] q_\rho \langle {\cal O}(z_1,\bar z_1)\cdots {\cal O}(z_n,\bar z_n)\rangle ,
}
again using that $P_{00}= qP$ and $P^\mu = \sum_{i=1}^n k_{i}^{\mu}$.  The above correlator simplifies to
\eq{
\sim \sum_{i=1}^n \frac{(\epsilon k_i)^2}{k k_i} qk \langle {\cal O}(z_1,\bar z_1)\cdots {\cal O}(z_n,\bar z_n)\rangle \sim \sum_{i=1}^n \frac{\bar z - \bar z_i }{z-z_i}  q k_i\langle {\cal O}(z_1,\bar z_1)\cdots {\cal O}(z_n,\bar z_n)\rangle,
}
where $qk=-1/2$ since $k$ is projective but $qk_i$ tracks the physical momentum of the hard particle in the $q$ direction.
As advertised, the right-hand side of this expression as precisely the leading Weinberg soft graviton factor \cite{weinberg} in our variables.

Comparing with \Eq{eq:STward}, we deduce that the holomorphic super-translation current is
\begin{equation} \label{eq:STcurrent}
j(z) = \partial_{\bar{z}} [\tilde t(z, \bar{z}), P_{00}].
\end{equation} 
Since the bulk-boundary propagator for $[\tilde t(z, \bar{z}), P_{00}]$ is a pure diffeomorphism, so too is the one for $j(z)$.  This suggests that there should again be a ``bulk'' topological description of the holomorphic 2D super-translation current, sensitive to the passage of hard particles.  

While this result is encouraging, there are several reasons why such a topological description of super-translations cannot be a straightforward CS theory.  First of all, from \Eq{eq:STcurrent}, we see that $j(z)$ is not a primary operator, as would be the case for the dual of a CS gauge field, and is instead descendant from a commutator of $\tilde{t}(z,\bar z)$. Relatedly, the global subgroup of super-translations, {\it i.e.}~ordinary translations, transform under the $SL(2, \mathbb C)$ Lorentz group, unlike  the global subgroup of a Kac-Moody algebra dual to a CS theory, which is $SL(2,\mathbb C)$ invariant. In any case, it would be very interesting to 
determine a bulk topological description for super-translations, if indeed one exists.

\section{Future Directions}
\label{sec:final}

A central result of this work is a recasting of 4D scattering amplitudes and their soft limits as correlators of a 2D CFT.  In particular, we showed that soft fields in 4D gauge theory and gravity have a description in terms of 3D CS theory en route to a mapping onto 2D conserved currents via  $\AdS_3/\CFT_2$. 
Remarkably, a number of physically significant aspects of 4D---soft theorems, asymptotic symmetries, and memory effects---are elegantly encoded as 2D Ward identities, their associated Kac-Moody and Virasoro symmetries, and 3D Aharonov-Bohm type effects. 
Of course, the results presented here are but a first step in exploring the possible implications of $\AdS/\CFT$ for flat space, and more generally, CS theories for describing soft gauge and gravitational phenomena.  Many questions remain, offering numerous avenues for future work that we now discuss.

First and foremost, we would like to better understand the role of unitarity in the 2D CFT, which cannot itself be unitary nor even a Wick rotation of a unitary CFT. 
Rather, since time is emergent, so too must be unitarity, which will then be non-manifest in the 2D description. On the other hand, starting from unitary 4D scattering amplitudes the 2D correlators must still somehow encode unitarity. However, what we really seek is some {\it independent principle} within the CFT guaranteeing 4D unitarity. 

Another open question relates to the role of 4D massive particles.   The foliation approach taken here is in principle consistent with such a generalization, but there will surely be new subtleties. Certainly with massive particles, the Weyl invariance used to simplify even the free particle analysis will be lost, and a more general complex set of scaling dimensions will arise. Relatedly, massive particles will not actually reach null infinity, but must ``sensed''  sub-asymptotically. 

More involved will be an extension of our results to loop level, where our foliation approach should apply.  With loops, it is likely that the CS description for soft gauge boson modes will have a 
level which depends on the infrared scale separating ``hard'' from ``soft'', due to the running of the gauge coupling.  An obvious exception is if the gauge coupling is at an infrared fixed point, in which case there may be a non-perturbative level free of infrared scale dependence. It would be interesting to understand whether the usual level quantization of CS theory implies that only 4D gauge theories with suitably quantized gauge couplings have a non-perturbative CS soft limit.

Furthermore, it was shown in \cite{ZviLoop} and \cite{SongHe} that the subleading soft theorems of  gauge theory and gravity are valid at tree level but are corrected at one-loop and higher.  Interestingly, these corrections appear to be critically tied to infrared divergences  \cite{ZviLoop2}.  This is naively quite disturbing because we saw that the subleading soft theorem for gravity is at the root of the Ward identities for the 2D stress tensor.\footnote{The Ward identities for holomorphic conserved currents on the other hand arise from the leading gauge and gravity soft theorems which are {\it not} loop-corrected, and are therefore unthreatened.} 
However, more carefully examined, there need be no actual conflict. The Ward identity for the 2D stress tensor is {\it related but not equal to} the subleading graviton soft theorem, which is corrected at one loop.  
 In fact, the complicated angular convolution in \Eq{eq:t_to_tildet} implies a highly non-trivial prescription for LSZ reduction that must be applied to the amplitude from the start.  It is possible that at loop level,  the 2D stress tensor 
continues to exist with some modified relationship to the Minkowski soft limit.
In any case, it is of utmost importance to study the robustness of our picture at loop level.  

A distinct but related question is to what extent the subleading soft theorems for gauge theory and the subsubleading soft theorems for gravity---which are known to be universal at tree level---might arise within the structure of the $\CFT_2$. For example, from the CFT perspective, new {\it non-conserved} vector currents should robustly arise from taking the conserved limit of non-conserved tensor operators \cite{maldacenaLimit}, which are AdS/CFT dual to the KK ``graviphoton'' of the effective compactification implied by the soft limit.

The OPE is a central feature of any CFT, which in the present context corresponds to the  structure of 4D collinear singularities. This suggests that the CFT structure may facilitate  some constructive method for building scattering amplitudes from collinear data.  This is reminiscent of the BCFW recursion relations, which when reduced down to three-particle amplitudes effectively does this.  On the other hand, the importance of self-dual configurations and the appearance of natural reference spinors $\eta$ and $\bar \eta$ throughout the discussion might naturally connect with CSW rule constructions for scattering amplitudes.  The focus on soft limits and collinear singularities also suggests connections with soft-collinear effective theory  \cite{stewartLectures,stewartSoft}, which may well be important for a loop-level formulation of asymptotic symmetries and the ideas presented in this paper.

There is also the question of whether our results can shed new light on the information paradox.  As proposed in \cite{softHair}, soft ``hair'' could offer an intriguing caveat to the usual picture of black hole information loss.  Nevertheless, stated purely in terms of soft radiation and gauge and gravitational memories, it is unclear how such a classical effect can resolve the paradox.  On the other hand, our results connect these effects to Aharonov-Bohm effects on the celestial sphere, which may offer a more quantum mechanical approach to this problem.
Also deserving of further study is our toy model for black hole horizons coming from the Rindler horizon of Minkowski spacetime.  In our picture, the restriction to the Rindler region revealed an extension of the CFT structure onto the past and future boundary of the horizon---effectively 
the dS/CFT dual of the past and future wavefunction of the horizon. Here, the CFT gives a description of this horizon, extending the notion of asymptotic symmetries in its presence. It would be interesting if these features, especially those related to topological structure of memories, extended to real black holes in less symmetric spacetimes. 

Finally, it would be worthwhile to see if the foliation approach followed here can be applied to spacetimes other than $\Mink_4$, for example $\AdS_4$, to uncover new symmetries and topological features emerging in special limits.

\begin{center} 
 {\bf Acknowledgments}
 \end{center}
 \noindent 
C.C.~is supported by a Sloan Research Fellowship and a DOE Early Career Award under Grant No.~DE-SC0010255. A.D. and R.S. are
supported in part by the NSF under Grant No.~PHY-1315155 and by the Maryland Center for Fundamental Physics.  R.S. would also like to thank the Gordon and Betty Moore Foundation for the award of a Moore Distinguished Scholar Fellowship to visit Caltech, as well as the hospitality of the Walter Burke Institute for Theoretical Physics, where a substantial part of this work was completed.  The authors are grateful to Nima Arkani-Hamed, Ricardo Caldeira Costa, Liam Fitzpatrick, Ted Jacobson, Dan Kapec, Jared Kaplan, Juan Maldacena, Ira Rothstein, and Anthony Speranza for useful discussions and comments.

\bibliographystyle{utphys} 
\bibliography{mink4cft2refs.bib} 
\end{document}